\documentclass[12pt]{article}

\usepackage{amssymb,verbatim,amsmath}

 \numberwithin{equation}{section}

\newtheorem{theorem}{Theorem}[section]

\newtheorem{remark}[theorem]{Remark}

\newtheorem{ex}{Example}[section]

\newtheorem{ass}{Assumption}[section]

\numberwithin{equation}{section}

\begin{document}

\newcommand{\ddt}{\partial \over \partial t}
\newcommand{\ddx}{\partial \over \partial x}
\newcommand{\ddy}{\partial \over \partial y}
\newcommand{\ddz}{\partial \over \partial z}
\newcommand{\ddr}{\partial \over \partial r}
\newcommand{\ddtheta}{\partial \over \partial \theta}
\newcommand{\dds}{\partial \over \partial s}

\newcommand{\ddphi}{\partial \over \partial  \varphi}

\newcommand{\ddu}{\partial \over \partial u}
\newcommand{\ddv}{\partial \over \partial v}
\newcommand{\ddw}{\partial \over \partial w}

\newcommand{\ddrho}{\partial \over \partial \rho }
\newcommand{\ddp}{\partial \over \partial p}

\newcommand{\ddEy}{\partial \over \partial E^y}
\newcommand{\ddEz}{\partial \over \partial E^z }
\newcommand{\ddHy}{\partial \over \partial H^y}
\newcommand{\ddHz}{\partial \over \partial H^z }

\newcommand{\ddEtheta}{\partial \over \partial E^{\theta}  }
\newcommand{\ddHtheta}{\partial \over \partial H^{\theta} }

\newcommand{\ddtbar}{\partial \over \partial \bar{t}}
\newcommand{\ddxbar}{\partial \over \partial \bar{x}}


\begin{center}
{\Large {\bf One-dimensional MHD flows with cylindrical  symmetry:   
Lie symmetries and conservation laws }}
\end{center}

\bigskip

\begin{center}
{\large Vladimir A.  Dorodnitsyn}$^{1,a}$, {\large  Evgeniy  I. Kaptsov}$^{2,b}$, \\
\smallskip
{\large Roman V. Kozlov}$^{3,c}$ , {\large Sergey  V. Meleshko}$^{2,d}$
\end{center}

\bigskip

\begin{tabbing}
$^1$ \= Keldysh Institute of Applied Mathematics, Russian Academy of Science, \\
\> Miusskaya Pl.~4, Moscow, 125047, Russia;\\
$^2$ \= School of Mathematics, Institute of Science, \\
\> Suranaree University of Technology, Nakhon Ratchasima, 30000, Thailand;\\
$^3$ \= Department of Business and Management Science, Norwegian School of Economics, \\
\> Helleveien 30, 5045, Bergen, Norway;
\\
\\
$^a$ \= Dorodnitsyn@Keldysh.ru \\
$^b$ \= evgkaptsov@math.sut.ac.th  \\
$^c$ \= Roman.Kozlov@nhh.no  \\
$^d$ \= sergey@math.sut.ac.th 
\end{tabbing}

\bigskip

\begin{center}
{\bf Abstract}
\end{center}
\begin {quotation}
A recent paper considered symmetries and conservation laws 
of the plane one-dimensional flows for magnetohydrodynamics 
in the mass Lagrangian  coordinates.  
This  paper analyses the one-dimensional magnetohydrodynamics flows 
with cylindrical symmetry
in the mass Lagrangian  coordinates. 
The medium is assumed inviscid and thermally non-conducting. 
It is modeled by a polytropic gas.  
Symmetries and conservation laws  are found. 
The cases of finite and infinite electric conductivity 
need to be analyzed separately.   
For finite electric conductivity $ \sigma ( \rho , p)  $  
we perform Lie group classification, 
which identifies  $ \sigma ( \rho , p)  $  cases with additional symmetries.  
The conservation laws are found by direct computation. 
For cases with infinite electric conductivity 
variational formulations of the equations are considered. 
Lie group classifications are obtained with the entropy treated as 
an arbitrary element. 
A variational  formulation allows to use the Noether theorem 
for computation of conservation laws. 
The conservation laws obtained for the variational equations  
are also presented in the original (physical)  variables. 
\end{quotation}

\bigskip
\bigskip


conservation laws; 

Noether theorem; 

Euler--Lagrange equations

\section{Introduction}

\label{Introduction}

Magnetohydrodynamics (MHD)  considers motion of electrically
conducting fluids subjected to electromagnetic forces. It models
different phenomena in  plasma physics, astrophysics and fluid
metals.

The recent paper~\cite{DKKMM2021}
was devoted to symmetries and  conservation laws
of the plane one-dimensional flows for  magnetohydrodynamics (MHD)
in the mass Lagrangian coordinates.
We refer to~\cite{DKKMM2021} and references therein
for a brief survey of the field  and
motivation to employ
Lie symmetry analysis~\cite{bk:Ovsiannikov1978, bk:Ibragimov1985, bk:Olver, bk:Bluman1989}.

The research concerning with MHD in cylindrical geometry
is devoted to exact and approximate solutions
as well as
to efficient computational approaches
for solving practical problems~\cite{Popov1971, Dorodnitsyn1973,
Tsui2005, lock_mestel_2008,  Pandey2008,
Suzuki2019}.
Particular attention is devoted to the analysis of
cylindrical shock waves~\cite{Arora2014, Nath2018, Devi2019, Chauhan2020,
Nath2020a,  Nath2020b, Singh2020,  
Nath2021, Nath2021a, Nath2021b, Nath2021,
 Singh2021, Singh2022}.

In the present paper we analyze one-dimensional MHD flows with
cylindrical symmetry. We consider Lie group symmetry properties and
conservation laws. We recall that Lie symmetries can  be used to
find particular solutions of differential equations, namely, group
invariant solutions. Conservation laws reflect important properties
of differential equations. Their preservation in numerical schemes
can be crucial for qualitatively  correct numerical
computations~\cite{Samarskii2001, Hairer2006, Furihata2011}.

Similarly to~\cite{DKKMM2021} the equations are considered in the
mass Lagrangian coordinates. The medium is assumed inviscid and
thermally non-conducting. It is modeled by a polytropic gas. Two
physically different cases of the electric conductivity, namely
finite and infinite conductivities, are examined. If electric
conductivity is infinite, magnetic force lines are "frozen" in mass
of liquid.

For the case of infinite electric conductivity
the MHD equations describing flows with cylindrical symmetry
can be brought into a variational form.
If differential equations have a variational formulation,
i.e. they are Euler--Lagrange equations  for some Lagrangian function,
one can employ the Noether theorem~\cite{bk:Noether1918}.
This theorem provides conservation laws
for those symmetries of the differential equations
which  are also variational or  divergence symmetries of the Lagrangian.
Other methods to obtain conservation laws
as well as other conserved quantities of MHD equations are reviewed  in~\cite{bk:Webb2018}.

The paper is organized as  follows.
The Noether theorem is briefly reviewed in the next Section.
In Section~\ref{Magnetohydrodynamics}
we start with general MHD equations,
restrict them to one-dimensional flows with cylindrical symmetry
and transfer the equations from Eulerian coordinates
into the mass Lagrangian coordinates.
We consider the case of finite electric conductivity
and present symmetries and conservation laws
in Sections~\ref{Symmetries_finite} and~\ref{Conservation_finite}, respectively.
Sections~\ref{Symmetries_infinite} and~\ref{Conservation_infinite}
provide symmetries and conservation laws
for the infinite electric conductivity.
Final Section~\ref{Concluding}
summarizes the results of the paper.
Several technical details are placed in the Appendices.

\section{Background results}

\label{Basic_theory}

In the following sections we examine the 
equations describing  one-dimensional MHD flows 
with cylindrical symmetry 
(in the mass Lagrangian coordinates)  
for admitted Lie point symmetries and conservation laws.  
To find conservation laws we employ 
either direct computation 
or the Noether theorem.

\subsection{Lie group classifications}

Lie group classification provides all Lie groups 
which are admitted by a system of 
differential equations~\cite{bk:Ovsiannikov1978, bk:Ibragimov1985, bk:Olver}. 
Lie groups are presented by their generators, 
often called symmetries.  
When the considered system has arbitrary elements 
(constants or functions), 
the classification specifies Lie groups admitted in the generic case, 
i.e. for all values of the arbitrary elements, 
and additional Lie groups 
admitted for special cases of the arbitrary elements. 
The set of generators  admitted in the generic case 
is called the kernel of the admitted Lie algebras.   
Lie group classifications are carried out with respect to equivalence transformations. 
Such transformations preserve the structure of the equations, 
however they can change the arbitrary elements.

\subsection{Noether theorem}

If equations have variational  structure, 
symmetries can be used to find conservation laws 
with the help of the Noether theorem~\cite{bk:Noether1918} 
(see also~\cite{bk:Ovsiannikov1978, bk:Ibragimov1985, bk:Olver, bk:Bluman1989}). 
In this point we provide a  simple version of the Noether theorem, 
which is specified for Section~\ref{Conservation_infinite}. 
We restrict ourselves to the case of two independent variables $(t,s)$ 
and first-order Lagrangians 
\begin{equation}     \label{Lagrangian}
  {L}={L}( t, s, \varphi   ,   \varphi   _t, \varphi   _s )  ,
\qquad
\varphi  =  ( \varphi ^1 , \ldots  ,      \varphi ^m  )  .
\end{equation}
This Lagrangian function generates the Euler--Lagrangian equations  
\begin{equation}     \label{EL_equations}
\frac{\delta{L}}{\delta \varphi   ^i  }
= { \partial  L \over \partial \varphi   ^i   }
-  D_t   \left( { \partial  L \over \partial  \varphi   ^i  _t }     \right)
-  D_s   \left( { \partial  L \over \partial  \varphi   ^i  _s }   \right)
 =    0  ,
\qquad
i = 1, \ldots , m ,
\end{equation}
which are second-order PDEs. 
Here $ D_t $ and $ D_s $  are total differentiation operators 
for time $t$  and  spacial variable $ s $.  
The equations define the variational operators $ \frac{\delta{L}}{\delta \varphi   ^i  }$.

Lie point symmetries are given by generators of the form 
\begin{equation}      \label{Symmetry}
X =\xi^{t} (t,s,  \varphi   )  \frac{\partial}{\partial t}
+\xi^{s} (t,s, \varphi   ) \frac{\partial}{\partial s}
+\eta^{i} (t,s,\varphi  ) \frac{\partial}{\partial  \varphi   ^i }   .
\end{equation}
The generators are assumed to be prolonged {to}  all variables 
involved in  the Euler--Lagrange  equations~(\ref{EL_equations}). 
The standard prolongation formulas can be found 
in~\cite{bk:Ovsiannikov1978, bk:Ibragimov1985, bk:Olver, bk:Bluman1989}.

The Noether theorem is grounded on two types of identities.    
The first identity is known as 
{\it the Noether identity}~\cite{bk:Noether1918, bk:Ibragimov1985}
\begin{equation}      \label{Noether_identity}
X  {L}+{L}(D_{t} \xi^{t}+D_{s}\xi^{s})
= ( \eta^{i}  - \xi ^t  \varphi   ^i _t  -  \xi ^s  \varphi   ^i _s  )
\frac{\delta{L}}{\delta \varphi   ^i  }
+ D_t ( N^t L ) +  D_s ( N^s L )    . 
\end{equation}
The operators  $ N^{t} $ and  $ N^{s} $  defined as 
\begin{equation}     \label{Noether_operators}
N^{t}
=\xi  ^{t}
+  (  \eta^{i} - \xi^{t}    \varphi ^i  _{t} - \xi ^{s}   \varphi ^i  _{s}  )
\frac{\partial  }{\partial     \varphi ^i   _{t} }   , 
\qquad
N^{s}
=\xi  ^{s}
+  (  \eta^{i} - \xi^{t}    \varphi ^i  _{t} - \xi ^{s}   \varphi ^i  _{s}  )
\frac{\partial  }{\partial     \varphi ^i   _{s} }   
\end{equation}
are called  Noether operators.  
This identity links the invariance  of the Lagrangian 
(more precisely the  invariance of the elementary action) 
with the conservation  laws 
existing on the solutions of the Euler--Lagrange equations.

There are two possibilities for invariance of the Lagrangian. 
If we have 
\begin{equation*}
X  {L}+{L}(D_{t} \xi^{t}+D_{s}\xi^{s}) = 0 ,
\end{equation*}
operator $X$ is called a {\it variational } symmetry of the Lagrangian.  
If there holds 
\begin{equation*}
X  {L}+{L}(D_{t} \xi^{t}+D_{s}\xi^{s}) = D_{t}  B _{1}   +  D_{s}  B _{2}
\end{equation*}
with nontrivial functions $  B _{1} ( t , s , \varphi ) $ and $  B _{1} ( t , s , \varphi ) $, 
operator $X$ is called a {\it divergence } symmetry of the Lagrangian.

The second type identities~\cite{bk:DorodnitsynKozlov[2011], bk:Olver}   
link invariance of the   Lagrangian 
with the invariance of the Euler--Lagrange equations 
\begin{multline}
\frac{\delta}{\delta \varphi ^j }
\left(
 X  {L}
+ {L}(D_{t} \xi^{t}+D_{s}\xi^{s}) \right)
=
X \left( \frac{\delta L}{\delta  \varphi ^{j}} \right)
\\
+
\left(\frac{\partial\eta^{k}}{\partial \varphi  ^{j}}
-\frac{\partial\xi^{t}}{\partial \varphi  ^{j}}  \varphi  _{t}^{k}
-\frac{\partial\xi^{s}}{\partial \varphi  ^{j}}  \varphi  _{s}^{k}
+     D_{t} \xi^{t}+D_{s}\xi^{s}  \right)
\frac{\delta L}{\delta \varphi  ^{k}}   ,
\qquad
j=1,2,  \ldots ,m . 
\end{multline}

The Noether theorem, 
adopted to the equations which will be considered in this paper,  
can be stated as

\begin{theorem}~\cite{bk:Noether1918}
Let Lagrangian~(\ref{Lagrangian}) 
satisfy 
\begin{equation}
   X  {L}+{L}(D_{t} \xi^{t}+D_{s}\xi^{s})
=D_{t}  B _{1}+D_{s}  B _{2}  ,
\label{eq:Noether}
\end{equation}
where generator $X$ is given by~(\ref{Symmetry}) 
and $ B_i = B_i ( t,s,\varphi) $, $ i = 1, \ 2 $. 
Then the operator $X$ is a symmetry of the Euler--Lagrange equations~(\ref{EL_equations}) 
and the conservation law 
\begin{equation}
D_{t}  (   N^{t} L - B _{1}  )
+
D_{s}   (   N^{s} L - B _{2} )
= 0  
\end{equation}
holds on the solutions of the Euler--Lagrange equations. 
\end{theorem}

\section{Magnetohydrodynamics equations and 
{one-}\\dimensional flows with cylindrical symmetry}

\label{Magnetohydrodynamics}

\subsection{MHD equations in three-dimensional space}

The magnetohydrodynamics are usually given in Cartesian coordinates 
$ ( x,y,z) $~\cite{bk:KulikovskiiLubimov1965, bk:SamarskyPopov_book[1992], bk:Landau_Electrodynamics}
(see also~\cite{Davidson2001,  Brushlinskii, Galtier2016,    bk:Webb2018}). 
In the dimensionless system  they are 
\begin{subequations}       \label{3D_system}
\begin{gather}
 \rho _t  +    \mbox{div} (  \rho   \textbf{u} ) =0 ,
\\
\textbf{u}  _t +    (  \textbf{u}  \cdot    \nabla  ) \textbf{u}  =
-  { 1 \over \rho }     \nabla   p
+   {    [  \textbf{i}  \times  \textbf{H} ]     \over  \rho  } ,
\\
 \varepsilon   _ t    +     (  \textbf{u}  \cdot    \nabla  ) \varepsilon   =
 -  { p \over \rho }   \   \mbox{div}   \  \textbf{u}
 +     { 1 \over \rho }    \   (  \textbf{i}    \cdot   \textbf{E}  )   ,
 \label{3D_system_energy}
\\
\textbf{H} _t
=    \mbox{rot}  \  [   \textbf{u}   \times   \textbf{H} ]
-     \mbox{rot} \   \textbf{E} ,
\qquad
\mbox{div} \ \textbf{H} = 0 ,
\\
\textbf{i}  =  \sigma  \textbf{E}  =   \mbox{rot}  \ \textbf{H}   .
 \label{relation}
\end{gather}
\end{subequations}
Here  the vector quantities, 
namely   
the velocity $ \textbf{u}$, 
 the electric current density $ \textbf{i} $, 
the magnetic field  $ \textbf{H} $ 
and 
the electric field $ \textbf{E} $,  
have components 
\begin{equation*}
\textbf{u}=(u^x  , u^y  , u^z   ) , 
\qquad  
\textbf{i}=(i ^x , i ^y  , i ^z  ) , 
\qquad  
\textbf{H}=(H ^x , H ^y  , H ^z  ) , 
\qquad  
\textbf{E}=(E ^x , E ^y  , E ^z  )    . 
\end{equation*}
The gradient operator is 
\begin{equation*}
 \nabla = \left(   {\ddx} ,    {\ddy} ,  {\ddz}  \right)  . 
\end{equation*}
The equations also contain 
the density $\rho $, 
the pressure  $p$  
and the internal  energy per unit volume  $  \varepsilon $. 
The last equation  can be used  to express  
$ \textbf{i} $ and  $ \textbf{E} $ 
via $ \textbf{H} $ and  the electric conductivity $ \sigma ( \rho , p) {\not \equiv }  0 $. 
It permits to eliminate $ \textbf{i} $ and  $ \textbf{E} $ 
from the system.

The set of equations~(\ref{3D_system}) is incomplete. 
It needs to be supplemented by  an equation of {\it state } 
\begin{equation*}
  \varepsilon = \varepsilon  (\rho , p) , 
\end{equation*}
which characterizes the considered medium. 
We model the   medium  as a gas which is 
ideal~\cite{bk:Chernyi_gas, bk:Ovsyannikov[2003]}
(see also~\cite{Landau, Chorin, Toro}), 
i.e.  
\begin{equation}   \label{ideal}
p = \rho R T,
\end{equation}
where $R$  is  the specific gas constant  and $T $ is the temperature, 
and polypropic, i.e. 
the internal energy is linear in temperature     
\begin{equation}      \label{EpsCvT}
\varepsilon (T) = c_{v} T  = { R T \over \gamma -1} . 
\end{equation}
Here $ c_v  $ is the specific heat capacity for constant volume and 
\begin{equation*} \label{gammaRCv}
\gamma = 1 + \frac{R}{c_{v}}  > 1
\end{equation*}
is the polytropic constant.

Equations~(\ref{ideal}) and~(\ref{EpsCvT}) allow to eliminate the temperature 
that provides the following equation of state   
\begin{equation} \label{InternalEnergyRel}
\varepsilon = \frac{1}{\gamma - 1} \frac{p}{\rho}.
\end{equation}

It is useful to consider  the entropy  $ \tilde{S} $, 
which relates the pressure and the density as 
\begin{equation}   \label{function_S}
p =  S  \rho^{\gamma} ,
\qquad
S = e^{(\tilde{S}-\tilde{S}_{0})/c_{v}} ,
\end{equation}
where $ \tilde{S}  _0$ is constant.

We remark that there are several ways to present  
the energy equation~(\ref{3D_system_energy}). 
It can be chosen  as the equation   
\begin{equation}      \label{equation_p_Euler}
p  _ t    +     (  \textbf{u}  \cdot    \nabla  )  p
+ \gamma    p     \   \mbox{div}  \   \textbf{u}
=    ( \gamma  -1 )      (  \textbf{i}    \cdot    \textbf{E}  )
\end{equation}
for  the pressure 
or as the equation 
\begin{equation}   \label{equation_S_Euler}
 S _ t    +     (  \textbf{u}  \cdot    \nabla  ) S    =
      {    \gamma  -1    \over \rho   ^{\gamma} }    \   (  \textbf{i}     \cdot   \textbf{E}  )   
\end{equation}
 for the entropy function  $ S $.

\subsection{Equations for one-dimensional flows 
with cylindrical symmetry  in  Eulerian coordinates}

In the  cylindrical coordinates   $ ( r, \theta, z)   $, 
which are related to Cartesian coordinates  $ ( x, y, z)   $ by 
\begin{equation*}
x = r  \cos \theta , 
\qquad
y  = r  \sin  \theta  , 
\end{equation*}
we switch to the corresponding components of the vector quantities
\begin{equation*}
{\bf u} 
=   ( u, v, w) , 
\qquad 
{\bf i} = ( i^r  , i ^{\theta} , i^z) , 
\qquad 
{\bf H} =  ( H^r , H^{\theta} , H^z)  , 
\qquad 
{\bf E} = ( E^r  , E^{\theta} , E^z)  .
\end{equation*}
For a vector $ (  a^x, a^y , a^z ) $  in Cartesian coordinates 
the  components in the cylindrical coordinates  $ (  a^r, a^{\theta}  , a^z ) $ 
are given by 
\begin{equation*}
a^ r = a^x \cos \theta  + a^y   \sin  \theta , 
\qquad
a^ {\theta}  = a^y \cos \theta  -  a^x   \sin  \theta  . 
\end{equation*}
The last component $ a^ z $ remains unchanged.

In the case of one-dimensional flows with cylindrical symmetry  
all dependent variables are functions of only two independent variables: 
time  $t$ and radius  $r$. 
In this case the equations 
\begin{subequations}
\begin{gather}
H^r   _t = 0   , 
\\
\mbox{div} \    \mbox{\bf H}   
= {  1 \over r }  {\partial   \over \partial r }   ( r H^r )   
=  0 
\end{gather}
\end{subequations}
provide
\begin{equation}    \label{A_value}
H^r   = { A \over r}  , 
\qquad 
A = \mbox{const} .
\end{equation}
We also obtain $ i^r = E^r = 0 $. 
The MHD system~(\ref{3D_system}) takes the following reduced form 
\begin{subequations}     \label{Euler_system} 
\begin{gather}
\rho _ t + {  1 \over  r }  ( r \rho   u ) _r  = 0  , 
 \label{Euler_system_rho} 
\\
\rho 
\left( 
u _ t + u u_r  -  { v^2 \over  r }  
\right)
= 
-     p_r 
-   { H  ^{\theta}  \over r }   \left(   r  H  ^{\theta}    \right) _r 
-   H^z  H  ^z _ r   , 
 \label{Euler_system_u} 
\\
\rho 
\left( 
v _ t + u v_r  +   { u v \over  r }   
\right) 
=   {  H^r  \over r  }  
\left(  
  r H  ^{\theta} 
\right)  _r  , 
 \label{Euler_system_v} 
\\
\rho (  w _ t + u w_r  )
=   {  H^r  }    H^z  _r     , 
 \label{Euler_system_w} 
\\
p_t   + u p_r + { \gamma   p \over r}   ( r u ) _r  
= (\gamma -1 ) \sigma  ( (  E  ^{\theta} ) ^2  +  ( E  ^z  ) ^2  )  , 
 \label{Euler_system_p} 
\\
H ^{\theta} _ t   
=    ( v H^r   - u H   ^{\theta}  + E^z )  _r , 
 \label{Euler_system_H_theta} 
\\
H^z _t  
= {  1 \over r }  \left( r ( w H^r    -   u H^z   -   E  ^{\theta}  )  \right)  _r  , 
 \label{Euler_system_H_z} 
\\
 \sigma   E ^{\theta} =  - H^z _ r  , 
\qquad 
\sigma   E ^z  =    {  1 \over  r }    (  r  H  ^{\theta}  )   _ r   . 
 \label{Euler_system_relations} 
\end{gather}
\end{subequations}
Here we eliminated   components of ${\bf i}$ from the system. 
The last two equations can be used to eliminate 
$ E ^{\theta} $ and $ E ^z  $.   
The equation for the energy~(\ref{3D_system_energy}) 
is replaced by the pressure equation~(\ref{equation_p_Euler}).

\begin{remark}
The case $ A \neq 0 $  can be considered only in a domain 
which does not include $z$-axis, 
i.e. the domain cannot have points with $r=0$.   
\end{remark}

\subsection{Equations for one-dimensional flows 
with cylindrical symmetry  in mass Lagrangian coordinates}

The detailed description how to introduce the mass Lagrangian  coordinates 
is given in Appendix~\ref{Appendix_Lagrangian_variables}. 
Briefly, the  mass Lagrangian   coordinates are introduced by the formulas 
\begin{equation}         \label{m_L_coordinates}
r =  \varphi (t, s) , 
\qquad 
\theta = \eta + \psi ( t , s ) , 
\qquad 
z  = \zeta  + \chi (t,s) , 
\end{equation}
where the functions $\varphi$,  $\psi $ and   $\chi $ solve the equations 
\begin{subequations}              \label{m_L_coordinates_details}
\begin{gather}
\label{m_L_coordinates_1}   
\varphi_{t} (t, s) =  u(t, \varphi(t, s) ), 
\qquad 
{ \varphi_{s} (t, s)} = {  1 \over    \varphi (t, s)   \rho(t, \varphi(t, s)) } ; 
\\
   \label{m_L_coordinates_2}   
\psi_{t} (t, s) = {  v (t, \varphi(t, s) ) \over  \varphi (t, s)   } ; 
\\
   \label{m_L_coordinates_3}   
\chi _{t} (t, s) =  w (t, \varphi(t, s) )   . 
\end{gather}
\end{subequations}

In the  mass Lagrangian coordinates $ (t,s)$  we rewrite the system~(\ref{Euler_system}) as 
\begin{subequations}    \label{Lagrange_system} 
\begin{gather}  
\rho_t = -\rho^2 (r u) _{{s}},    
  \label{Lagrange_system_rho} 
\\
 u_t   -  { v^2 \over  r }  = - r p_{{s}}
-   { 1  \over 2  r  }       \left(   r^2  ( H^{\theta} )^2     \right)  _s 
-   { r  \over 2 }       \left(    ( H^z )^2     \right)  _s  ,
\qquad
r_t = u ,  
 \label{Lagrange_system_u} 
\\
 v _t   +   {  u v  \over  r }  
=  { H^r  }     \left(   r   H^{\theta}    \right)  _s , 
\qquad 
{\theta} _t  = {  v \over r }    ,  
 \label{Lagrange_system_v} 
\\
 w_t      
=  r { H^r }         H^{z}     _s , 
\qquad 
z_t =  w , 
 \label{Lagrange_system_w} 
\\
   p_t  
= - \gamma  \rho p ( r u )_{{s}}
+  (\gamma -1 ) \sigma  ( (  E  ^{\theta} ) ^2  +  ( E  ^z  ) ^2  )  , 
  \label{Lagrange_system_p} 
\\     
  H^{\theta}  _t  
= 
r \rho (   ( v H^r   + E^z )  _s    -  H^{\theta} u_s )  , 
 \label{Lagrange_system_H_theta} 
\\ 
  H^z _t 
=  \rho  (      ( r w H^r   -  r E  ^{\theta})  _s      -   H^z  (r u )_s   )     , 
  \label{Lagrange_system_H_z} 
\\
 \sigma   E ^{\theta} =  -  r \rho H^z _ s  , 
\qquad 
 \sigma   E ^z  =    \rho    (  r  H  ^{\theta}  )   _ s   . 
 \label{Lagrange_system_relations} 
\end{gather}
\end{subequations}
We stress that the system 
uses Lagrangian time differentiation, 
which is taken along the trajectories of motion. 
For this reason the system includes the dynamical  equations for $r$, $\theta$ and $z$.  
The relation between Lagrangian and Eulerian  
total differentiation operators will be addressed 
in Remark~\ref{L_E_relation}.

A particular case of the system~(\ref{Lagrange_system})  with  $H^z  \equiv 0$ 
was considered  in Samarskii~\cite{bk:SamarskyPopov_book[1992]}. 
The assumption $H^z  \equiv 0$  leads to $ w_t = 0 $  and $ E ^{\theta} = 0 $ 
that provides a substantial simplification of the system.

The Eulerian spatial coordinate $r$ is a nonlocal variable 
in the mass Lagrangian coordinates $(t,s)$. 
It is defined by the equations    
\begin{equation}    \label{nonlocal_r} 
r_t = u ,  
\qquad 
r_s = { 1 \over r \rho }  . 
\end{equation}
We note that in the mass Lagrangian coordinates  we also have 
\begin{equation}      \label{nonlocal_theta_z} 
{\theta} _t  = {  v \over r }    ,  
\qquad
z_t =  w . 
\end{equation}

\begin{remark} 
Using~(\ref{Lagrange_system_rho}), 
it is possible to rewrite equation~(\ref{Lagrange_system_H_theta})  as 
 \begin{equation*}     
  \left(   
H^{\theta}   \over  \rho 
\right)      _t
   =    
r   \left( v H^r  + E^z  \right)  _s  
+ { u  H^{\theta}   \over  r  \rho   }  
 \end{equation*}
 that can be further rewritten as conservation law  
\begin{equation}         \label{Modification_H_theta}
\left(   
H^{\theta}   \over  r \rho   
\right)      _t
= 
\left( 
v H^r  + E^z  
\right)  _s  . 
\end{equation} 
Similarly,  equation~(\ref{Lagrange_system_rho})  permits  
to rewrite equation~(\ref{Lagrange_system_H_z})    as  conservation law  
\begin{equation}         \label{Modification_H_z}
  \left(   
H^z   \over  \rho 
\right)      _t
  =  
\left( 
r w H^r  - r E ^{\theta} 
\right) _s   . 
\end{equation}   
\end{remark}

\section{Lie group classification of equations~(\ref{Lagrange_system}) in the case of finite  conductivity}

\label{Symmetries_finite}

Here we start to consider different cases of system~(\ref{Lagrange_system}). 
It turns out (see equations~(\ref{classifying_equation_general_0})) 
that we need to consider two cases for the arbitrary constant $ A$: 
$A \neq 0 $ and  $A = 0 $. 
For each case we perform Lie group classification 
with  an arbitrary  function   $\sigma ( \rho , p) $.

Lie symmetry groups are given by the infinitesimal generators  
\begin{multline} \label{X0}
X = \xi^t \frac{\partial}{\partial{t}}
    + \xi^s \frac{\partial}{\partial{s}}
    + \eta^r \frac{\partial}{\partial{r}}
    + \eta^{\theta} \frac{\partial}{\partial{\theta}}
    + \eta^z \frac{\partial}{\partial{z}}
    + \eta^u \frac{\partial}{\partial{u}}
    + \eta^v \frac{\partial}{\partial{v}}
    + \eta^w \frac{\partial}{\partial{w}}
\\
+ \eta^\rho \frac{\partial}{\partial{\rho}}
+ \eta^p \frac{\partial}{\partial{p}}
    + \eta^{E^{\theta}} \frac{\partial}{\partial{E^{\theta}}}
    + \eta^{E^z} \frac{\partial}{\partial{E^z}}
    + \eta^{H^{\theta}} \frac{\partial}{\partial{H^{\theta}}}
    + \eta^{H^z} \frac{\partial}{\partial{H^z}} .
\end{multline}
The coefficients of this generator 
 $\xi^t$, $\xi^s$, $\eta^r$, ..., $\eta^{H^z}$ 
are functions of  variables 
$t$, $s$, ${r}$,  $ {\theta} $, $z$,  ${u}$, $v$, $w$, $\rho$, $p$,  
$E^{\theta}$, $E^z$,  $H^{\theta}$ and $H^z$, 
i.e. functions of all  independent and dependent variables.

To find the admitted Lie symmetries 
we employ the infinitesimal criterion of invariance~\cite{bk:Ovsiannikov1978, bk:Ibragimov1985,  bk:Olver}. 
It states that we prolong the generator 
to all derivatives involved in the system  of equations, 
apply the prolonged  generator to the equations 
and require the result to be identity on the solutions of the system. 
This can be presented as   
\begin{equation} \label{classifDetsysF}
\left.  X  ( \mathcal{F} )  \right|_{ \mathcal{F} = 0 }= 0,
\end{equation}
where $\mathcal{F} = 0 $ denotes the considered system.  
Splitting the left hand side of~(\ref{classifDetsysF}) 
with respect to derivatives, we obtain the determining equations, 
which provide the coefficients of the generator~(\ref{X0}).

Splitting the system of the  determining equations with respect to the first-order derivatives 
and performing simplifications, we arrive at the following {\it classifying}  system 
\begin{subequations}     \label{classifying_equation_general_0}
\begin{gather} 
2 (  a_3 - 2 a_4   +  a_5 ) \rho \sigma  _{\rho}   ( \rho , p ) 
+ 2 ( -  a_4   +  a _5 )  p  \sigma  _{p}     ( \rho , p )
=  ( a_3 - 2  a_4 ) \sigma   ( \rho , p )   , 
 \label{classifying_equation_general_1}
\\
A \eta ^{\theta}  _r  (s, r v) = 0,  
 \label{classifying_equation_general_2}
\\
A \eta^{ \theta} _v  (s, r v) = 0, 
 \label{classifying_equation_general_3}
\\
A   a _5 = 0 
 \label{classifying_equation_general_4}
\end{gather}
\end{subequations}
for the arbitrary elements  $  \sigma ( \rho , p ) $ and $ A$. 
The coefficients of the symmetry generator~(\ref{X0}) are given by 
\begin{equation*} 
\xi^t = a_3 t + a_1 , 
\qquad 
\xi^s = 2 ( a_3 - a_4 + a_5 )  s + a_2 , 
\end{equation*}
\begin{equation*} 
\eta^r = a_4  r , 
\qquad  
\eta^{\theta}  = f_1(s , r v   ) , 
\qquad 
\eta^z = f_2 (s) +  a _4 z + a_6 t   , 
\end{equation*}
\begin{equation*} 
\eta^u = ( -  a_3  + a _4  )  u , 
\qquad  
\eta^v = ( -  a_3  + a _4   )  v , 
\qquad  
\eta^w = ( -  a_3 + a _4   ) w + a_6 , 
\end{equation*}
\begin{equation*} 
\eta^{\rho} = 2 ( a_3  -  2 a _4 + a_5 )  \rho , 
\qquad  
\eta^p =  2  ( - a_4 + a_5 )  p , 
\end{equation*}
\begin{equation*} 
\eta^{E^{\theta}} =   ( -  a_3  + a _5  )    E^{\theta} , 
\qquad  
\eta^{E^z} =   ( -  a_3  + a _5  )   E^z , 
\end{equation*}
\begin{equation*} 
\eta^{H^{\theta}} =   ( -  a_4  + a _5  )       H^{\theta} , 
\qquad  
\eta^{H^z} =    ( -  a_4  + a _5  )   H^z , 
\end{equation*}
where the functions $ f_1 (s, r v ) $ and $ f_2 (s) $ are arbitrary.    
Here and in the rest of the paper the coefficients $ a_i $  take constant values.    
It follows that   
the symmetries of system~(\ref{Lagrange_system})  are expanded in the basis 
\begin{multline}  \label{extended_algebra_1}
Y_1 = {\ddt},
\qquad
Y_2 =  {\dds},
 \\
Y_3 = t   {\ddt} + 2 s   {\dds}
    - u  {\ddu} - v  {\ddv} - w  {\ddw}
    + 2\rho  {\ddrho}
    - E^{\theta}   { \partial \over \partial  E^{\theta} }   
 -  E^{z}   { \partial \over \partial  E^{z} }   ,
\\
Y_4
= - 2 s   {\dds}
+ r  {\ddr}   + z  {\ddz}
+ u  {\ddu} + v  {\ddv}  + w  {\ddw}
    - 4 \rho  {\ddrho}  - 2 p {\ddp}   
    -  H^{\theta}  {\ddHtheta} -  H^z  {\ddHz}, 
\\
Y_5 = 2 s   {\dds}
   + 2\rho  {\ddrho}
    + 2 p   {\ddp}
    + E^{\theta}   {\ddEtheta} +  E^z   {\ddEz}
    + H^{\theta}   {\ddHtheta} +  H^z  {\ddHz},
\\
Y_{6} =   t  {\ddz} +  {\ddw} , 
\qquad 
Y_7
=  f _1 (s, r v ) { \partial \over \partial \theta} ,  
\qquad 
Y_{8} = f_2 (s)  {\ddz} . 
\end{multline}

For further analysis we need to consider the cases 
$ A \neq 0 $  and  $ A = 0  $  separately.

\subsection{Case   $ A \neq 0$}

Now we develop Lie group classifications for system~(\ref{Lagrange_system}) 
with $ A \neq 0 $.  
We have arbitrary elements $ \sigma ( \rho , p )$  and $ A $.

Equivalence transformations can be used to change arbitrary elements 
without any changes of the structure of the equations~\cite{bk:Ovsiannikov1978}.  
The equivalence transformations  for system~(\ref{Lagrange_system}) 
with arbitrary $ \sigma ( \rho , p )$  and arbitrary $ A $ 
are provided 
in~(\ref{equivalence_2}), Appendix~\ref{Equivalence}.  
They allow scaling of the  function $ \sigma ( \rho , p )$  and the constant $ A $.

\subsubsection{Arbitrary  $ \sigma ( \rho , p) $}

The kernel of the admitted Lie algebras contains 
the generators of the group transformations  
admitted by the system for all choices of the  arbitrary elements.  
For system~(\ref{Lagrange_system}) 
with arbitrary function $ \sigma ( \rho , p )$  and arbitrary constant $ A $  
we obtain the kernel specified by generators 
\begin{equation}     \label{kern01}
X_1 = {\ddt},
\quad
X_2 =  {\dds},
\quad
X_{3} =  t  {\ddz} +  {\ddw}  ,
\quad
X_4
=  h_1 (s) { \partial \over \partial \theta} , 
\quad
X_{5} = h _2 (s)   {\ddz},
\end{equation}  
where $ h _1  (s) $  and  $ h _2  (s) $  are arbitrary functions.

\subsubsection{Special cases of $ \sigma ( \rho , p) $}

For particular cases of function $  \sigma ( \rho , p )$   
there can be  
additional symmetries of system~(\ref{Lagrange_system}). 
Such symmetries have the form 
\begin{equation} 
a_3 Y_3 + a_4 Y_4  , 
\end{equation} 
where $ Y_3  $ and $  Y_4 $ are generators given in~(\ref{extended_algebra_1}).   
The constant coefficients $  a_3 $  and $ a_4  $  
and the  function   $  \sigma  ( \rho , p ) $ must satisfy 
the  classifying  equation
\begin{equation}     \label{classifying_equation_with_A}
2 (  a_3 - 2 a_4 ) \rho \sigma  _{\rho}    ( \rho , p ) 
-     2  a_4  p  \sigma  _{p}    ( \rho , p ) 
=  ( a_3 - 2  a_4 ) \sigma  ( \rho , p )   , 
\end{equation}
which follows from~(\ref{classifying_equation_general_1}) 
and~(\ref{classifying_equation_general_4}).  
For analysis of the classifying equation we apply the classical approach 
proposed in~\cite{bk:Ovsiannikov1978}.

If   $  a_3 =  2 a_4 \neq 0 $, the classifying equation takes the form 
\begin{equation*} 
\sigma  _{p}    ( \rho , p )   =  0
\end{equation*}
and provides solutions   $ \sigma  = F ( \rho ) $, where $ F $  is an arbitrary function.

If   $  a_3 \neq   2 a_4 $, we rewrite the classifying equation as 
\begin{equation*} 
\rho \sigma  _{\rho}     ( \rho , p ) 
-  \alpha    p  \sigma  _{p}     ( \rho , p ) 
=    {  1  \over  2 }  \sigma   ( \rho , p )  , 
\qquad 
\alpha  =  {   a_4 \over  a_3 - 2  a_4 }  . 
\end{equation*}
It has solutions  $ \sigma  = \sqrt{\rho} F ( p \rho ^{ \alpha} ) $ 
with arbitrary function $F$.

We summarize the obtained results in Table~1.
The first column provides the dimension $\mbox{dim}   \  L $
of the Lie algebra.  
The second column gives the additional symmetries, 
i.e. the extensions of the kernel of the admitted Lie algebras~(\ref{kern01}). 
The third column presents  the functions  $\sigma ( \rho, p) $.

\begin{table}[ht]
\def\arraystretch{1.75}
\centering
\begin{tabular}{|c|l|l|}
\hline
$\mbox{dim} \  L $   & Extension of kernel~(\ref{kern01}) & $\sigma   ( \rho, p)   $ \\
\hline
$6$ &
$  X_{6} =  2  Y_3 +    Y_4 $
&   $  F(  \rho ) $  \\
   &
$  X_{6} = ( 1 + 2 \alpha )   Y_3 +   \alpha    Y_4 $
&   $ \sqrt{\rho}  F( p  \rho ^{\alpha} )$  \\
\hline
\end{tabular}
\label{tab:class_lagr_flat_H_neq_0}
\caption{Lie group extensions for $A  \neq 0$.}
\end{table}

\begin{remark}
Equation~(\ref{ideal}) can be used to present some   cases $\sigma  ( \rho, p)  $
in the form~$\tilde{\sigma}(T)$.  
Such presentations are of interest in  physical applications.  
For  the particular case  $  F(x) =  C {x} ^{- 1\over 2( \alpha +1 ) }$ 
the second case of the Table~1, 
namely  $ \sigma = \sqrt{\rho}  F( p  \rho ^{\alpha} )$, 
can be rewritten as  
$ \sigma =  \tilde{C} T  ^{- 1\over 2( \alpha +1 ) }$. 
\end{remark}

\subsection{Case   $ A = 0$}

\label{finite_sigma_A_0}

For the case  $ A = 0$ we get $ H^r \equiv 0$. 
The system~(\ref{Lagrange_system})  splits into the reduced system 
\begin{subequations}    \label{Lagrange_system2} 
\begin{gather}  
\rho_t = -\rho^2 (r u) _{{s}},    
  \label{Lagrange_system2_rho} 
\\
 u_t   -  { v^2 \over  r }  = - r p_{{s}}
-   { 1  \over 2  r  }       \left(   r^2  ( H^{\theta} )^2     \right)  _s 
-   { r  \over 2 }       \left(    ( H^z )^2     \right)  _s  ,
\qquad
r_t = u ,  
 \label{Lagrange_system2_u} 
\\
 v _t   +   {  u v  \over  r }  
= 0 , 
\qquad 
{\theta} _t  = {  v \over r }    ,  
 \label{Lagrange_system2_v} 
\\
   p_t  
= - \gamma  \rho p ( r u )_{{s}}
+  (\gamma -1 ) \sigma  ( (  E  ^{\theta} ) ^2  +  ( E  ^z  ) ^2  )  , 
  \label{Lagrange_system2_p} 
\\     
  H^{\theta}  _t  
= 
r \rho (     E^z  _s    -  H^{\theta} u_s )  , 
 \label{Lagrange_system2_H_theta} 
\\ 
  H^z _t 
=  -   \rho  (       (    r E  ^{\theta})  _s      +    H^z  (r u )_s   )     , 
  \label{Lagrange_system2_H_z} 
\\
 \sigma   E ^{\theta} =  -  r \rho H^z _ s  , 
\qquad 
 \sigma   E ^z  =    \rho    (  r  H  ^{\theta}  )   _ s   
 \label{Lagrange_system2_relations} 
\end{gather}
\end{subequations}
and the remaining equations  
\begin{equation}      
 w_t      
=  0 , 
\qquad 
z_t =  w  . 
 \label{Lagrange_system2_w} 
\end{equation}
The last two equations~(\ref{Lagrange_system2_w}) 
can be separated from the main system~(\ref{Lagrange_system2}). 
They are solved as 
\begin{equation}  \label{w_solved}
w = w_0 (s), 
\qquad 
z = w_0 (s) t  + z_0  ( s, \eta , \zeta)  .
\end{equation}
Further  simplification of the system~(\ref{Lagrange_system2}) 
is possible if we  restrict to radial motion, i.e. assume $ v  ( t,s) \equiv  0$.

In the rest of this point we consider the reduced system~(\ref{Lagrange_system2}).

\begin{remark} 
Using  equation~(\ref{Lagrange_system2_rho}), 
we can rewrite equation~(\ref{Lagrange_system2_H_theta})  as 
 \begin{equation*}     
  \left(   
H^{\theta}   \over  \rho 
\right)      _t
   =    
r   \left(  E^z \right) _s  
+ { u  H^{\theta}   \over  r  \rho   }  
 \end{equation*}
 that can be further simplified as 
\begin{equation}         \label{Modification2_H_theta}
\left(   
H^{\theta}   \over  r \rho   
\right)      _t
= 
\left( E^z  \right) _s  . 
\end{equation}
At the same  time, 
it is possible to rewrite equation~(\ref{Lagrange_system2_H_z})    as 
\begin{equation}         \label{Modification2_H_z}
  \left(   
H^z   \over  \rho 
\right)      _t
  =  
  -  \left( r E ^{\theta} \right) _s   . 
\end{equation}   
\end{remark}

The equivalence transformations for system~(\ref{Lagrange_system2}) 
are given in~(\ref{equivalence_4}), Appendix~\ref{Equivalence}.   
Symmetries of system~(\ref{Lagrange_system2})  
are expanded in the basis 
\begin{multline}  \label{extended_algebra_2}
Y_1 = {\ddt},
\qquad
Y_2 =  {\dds}, 
\\
Y_3 = t   {\ddt} + 2 s   {\dds}
    - u  {\ddu} - v  {\ddv} 
    + 2 \rho  {\ddrho}
    - E^{\theta}   { \partial \over \partial  E^{\theta} }   
 -  E^{z}   { \partial \over \partial  E^{z} }   ,
\\
Y_4
= - 2 s   {\dds}
+ r  {\ddr}   
+ u  {\ddu} + v  {\ddv}  
    - 4 \rho  {\ddrho}  -  2 p  {\ddp}
    -  H^{\theta}  {\ddHtheta}  -  H^z  {\ddHz},
\\
Y_5 = 2 s   {\dds}
    + 2\rho  {\ddrho}
    + 2 p   {\ddp}
    + E^{\theta}   {\ddEtheta} +  E^z   {\ddEz}
    + H^{\theta}   {\ddHtheta} +  H^z  {\ddHz},
\\
Y_6
=  f (s , r v )  { \partial \over \partial \theta} , 
\end{multline}
where $ f   (s , r v ) $ is an arbitrary function. 
Note that we can  consider this basis as truncation of basis~(\ref{extended_algebra_1}): 
variables $z$  and $w$ are excluded.

\subsubsection{Arbitrary  $ \sigma ( \rho , p) $}

For arbitrary function  $ \sigma ( \rho , p) $ 
system~(\ref{Lagrange_system2})
 admits the following generators 
\begin{equation}     \label{kern02}
X_1 = {\ddt},
\qquad
X_2 =  {\dds},
\qquad
X_3
=  h ( s, r v ) { \partial \over \partial \theta} , 
\end{equation}  
where   function  $  h ( s, r v )  $ is arbitrary. 
These generators form the kernel of the admitted Lie algebras.

\subsubsection{Special cases of $ \sigma ( \rho , p) $}

For particular cases of   $ \sigma ( \rho , p) $ 
there can  be additional symmetries of the form 
\begin{equation}     \label{extension_form} 
a_3 Y_3 +  a_4 Y_4 + a_5 Y_5  ,  
\end{equation}
where  $ Y_3  $,    $ Y_4  $ and  $ Y_5  $ are given  in~(\ref{extended_algebra_2}).  
For $ A = 0 $ the classifying system~(\ref{classifying_equation_general_0}) 
has  only one non-trivial equation, namely 
the classifying equation 
\begin{equation}     \label{classifying_equation_without_A}
2 (  a_3 - 2 a_4   + a_5 ) \rho \sigma  _{\rho}     ( \rho, p) 
+ 2 ( -     a_4   + a _5 )  p  \sigma  _{p}   ( \rho, p) 
=  ( a_3 - 2  a_4 ) \sigma  ( \rho, p)  .
\end{equation}

This equation allows to find  the cases of conductivity $ \sigma ( \rho , p) $ 
for which there can be symmetry extensions. 
We recall that coefficients $ a_i  $ take constant values 
and that 
the proportional sets of coefficients   $ ( a  _3, a _4,  a  _5 ) $ 
and    $ ( a ' _3, a ' _4,  a '  _5 ) $ provide the same symmetry~(\ref{extension_form}).   
Thus, we look for functions    $ \sigma ( \rho , p) $  
 for which  the coefficients space $ ( a  _3, a _4,  a  _5 ) $ 
is  one-dimensional, 
 two-dimensional 
or  
three-dimensional. 
We remark that this classification method  was suggested in~\cite{bk:Ovsiannikov1978}.

For convenient presentation of the admitted generators 
we start with choice  of a basis for the coefficients space.  
If the coefficients  $ ( a  _3, a _4,  a  _5 ) $  
form a one-dimensional vector space, 
 we can assume that the basis of this space 
has one of the following forms 
\begin{equation*} 
\{ ( 1, 0, 0) \} , 
\qquad
  \{ ( \alpha , 1, 0) \} , 
\qquad 
\{ ( \beta , \alpha , 1) \} ,
\end{equation*}
where  $ \alpha $  and $\beta $  are arbitrary constants.   
For all three {sub}cases  there exist  functions $ \sigma ( \rho , p) $ 
which solve~(\ref{classifying_equation_without_A}).  
These  functions $ \sigma ( \rho , p) $ and symmetry extensions are given  in Table~2.

When the coefficients  $ ( a  _3, a _4,  a  _5 ) $  
belong to a two-dimensional vector space, 
 we can take the basis of this space as one of the following 
\begin{equation*} 
\{ ( 1, 0, 0) , ( 0, 1, 0) \} , 
\qquad
  \{  ( 1, 0, 0) ,  ( 0, \alpha , 1) \} , 
\qquad 
\{( \alpha , 1, 0)  ,  ( \beta , 0 , 1) \} ,
\end{equation*}
where  $ \alpha $  and $\beta $  are arbitrary constants.    
All three cases are realized: 
for all three cases there are  functions $ \sigma ( \rho , p) $ 
solving~(\ref{classifying_equation_without_A}) for both elements of each basis case.   
We provide the results in Table~2.

If the coefficients  $ ( a  _3, a _4,  a  _5 ) $  
make up the whole three-dimensional space, 
 we can choose the basis as 
\begin{equation*} 
\{ ( 1, 0, 0) , ( 0, 1, 0) ,   ( 0, 0 , 1) \} . 
\end{equation*}
This case is not realized:   
the system of equations~(\ref{classifying_equation_without_A}) 
for all three elements of this basis gives $ \sigma ( \rho , p) \equiv 0$.

\begin{table}[ht]
\def\arraystretch{1.5}
\centering
\begin{tabular}{|c|l|l|}
\hline
$\mbox{dim}  \   L $
& Extension of kernel~(\ref{kern02})
& $\sigma  ( \rho , p)  $
\\
\hline
$ $
&  $ X_4 = Y_3 $
& 
 $ \sqrt{\rho}  F(p)$
\\
4  & $ X_4 = Y_4 + \alpha Y_3 $
&  
 $   \sqrt{\rho}  F  \left(  p   \rho    ^{1 \over  \alpha - 2 }  \right)$
\\
& $ X_4 = Y_5 + \alpha Y_4 + \beta Y_3 $
&  
 $\rho ^{  2 \alpha  - \beta  \over 2 ( 2 \alpha  - \beta -1 ) }     
 F \left( p   \rho ^{  1 -  \alpha  \over  2 \alpha  - \beta -1  }      \right)$
\\
\hline
&
$ X_4 = Y_3, 
\quad 
X_5 = Y_4 $
&   
$ C \sqrt{\rho}    $
\\
${5}$ 
&
$ X_4 =  Y_3, 
\quad 
X_5 = Y_5 + \alpha Y_4 $
&   
$  C   \sqrt{\rho}    p^{ 1 \over 2 (\alpha  -1 )}$
\\
& 
$  X_4 = Y_4 + \alpha Y_3 , 
\quad 
X_5 = Y_5 + \beta  Y_3 $
&   
$ C 
\rho ^{     \alpha +  \beta   -2  \over 2 (   \alpha  + \beta   - 1 )   } 
p  ^{    2   - \alpha   \over 2 (   \alpha +   \beta   - 1 )   }  $
 \\
\hline
\end{tabular}
\label{tab:class_lagr_flat_H_eq_0}
\caption{Lie group extensions for  $A = 0$.}
\end{table}

\begin{remark}
There are particular cases of  $ F(p)  $ for four-dimensional algebras 
and particular cases  of  $ \alpha$ for five-dimensional algebras  
when the electric conductivity $\sigma  ( \rho , p)  $  
can be  rewritten as a function of the temperature $T$:

\begin{itemize}

\item

Case  $ X_4 = Y_3 $

For   $    F(x) = C x^{ -1 /2 }  $ 
we obtain 
$ \tilde{\sigma}(T)    = \tilde{C}  T^{  -1 /2 }   $.

\item

Case  $ X_4 = Y_4 + \alpha Y_3 $

Function  $    F(x) = C  x ^{  { 2 - \alpha  \over 2  (   \alpha - 1 )  } }$
leads to 
$   \tilde{\sigma}(T)    = \tilde{C}     T  ^{   { 2 - \alpha \over 2  (   \alpha - 1 ) }    } $.

\item

Case $ X_4 = Y_5 + \alpha Y_4 + \beta Y_3 $

For    $    F(x ) =  C    x ^{ \beta  -  2 \alpha  \over 2  (   \alpha -  \beta  ) } $
we get 
$    \tilde{\sigma}(T)   = \tilde{C}   T  ^{ \beta  -  2 \alpha  \over 2  (   \alpha -  \beta  ) }     $.

\item

Case    $  X_4 = Y_3 $,   $  X_5 = Y_5 + \alpha  Y_4 $

If  $ \alpha = 0 $, 
we obtain 
$    \tilde{\sigma}(T)   = \tilde{C}    T^{   -1 /2 }   $.

\item

Case    $  X_4 = Y_4 + \alpha   Y_3 $,   $  X_5 = Y_5 + \beta   Y_3 $

With  $ \beta  =0 $
we get
$    \tilde{\sigma}(T)   = \tilde{C}   T  ^{ 2 - \alpha   \over 2  (   \alpha   - 1  ) }      $.

\end{itemize}
\end{remark}


\section{Conservation laws  for finite  conductivity}

\label{Conservation_finite}

In the case of two independent variables $ (t,s) $ 
conservation laws have the form 
\begin{equation} \label{CLgenform}
D_t ^L ( {T}^t ) + D_s(  {T}^s )   
= 0 .
\end{equation}
The conservation laws  should hold on the solutions of the considered PDEs. 
For MHD equations~(\ref{Lagrange_system}) 
the densities     $ {T}^t  $  and     $ {T}^s  $  are functions of all independent 
and  dependent variables
$ ( t, s, r, \theta , z , u, v, w,  \rho, p,  E^{\theta} ,  E^z,  H^{\theta}  ,  H^z  )  $.

In this paper we use direct computations 
to find conservation laws in the case of finite electric conductivity 
and 
employ the Noether theorem  
to obtain conservation laws in the case of infinite electric conductivity. 
We refer to~\cite{RNaz2008} for a review 
of different methods which can be used to find conservation laws.

\begin{remark}    \label{L_E_relation}
It is possible to rewrite conservation laws~(\ref{CLgenform}), 
which are found in the Lagrangian coordinates, 
as   conservation laws  
\begin{equation}
D_{t} ^E  (  { ^{e} T^{t} }  )  +  D_{r}  (  { ^{e} T^{r} }  ) = 0 
\end{equation}
in the Eulerian coordinates.

We recall that 
for one-dimensional MHD flows with cylindrical symmetry 
the total differentiation operators  
in the Lagrangian coordinates  
$ D_{t} ^L  $ and $  D_{s} $  
are related to the total differentiation operators  
in the Eulerian coordinates   
$ D_{t} ^E  $ and $  D_{r} $ 
as follows 
\begin{equation}
D_{t} ^L
= D_{t}  ^E+ u D_{r}   ,
\qquad
D_{s}
=\frac{1}{r \rho}D_{r} .
\end{equation}

The  densities of the conservation laws 
in the Eulerian and Lagrangian  coordinates  are related as 
\begin{equation}     \label{transformation_rule}
^{e}T^{t}=r \rho T^{t}  ,
\qquad
{}^{e}T^{r}= r \rho u T^{t}+T^{s}  .
\end{equation}
This result follows from
\begin{equation*}
D_{t} ^L ( T^{t} ) +D_{s}  ( T^{s} ) 
=  
{  1 \over r \rho }   \left(    D_{t}  ^E (r \rho T^{t})   
 +  
D_{r}(    r  \rho u T^{t}+T^{s} )    \right)
\end{equation*}
that can be verified by direct computation.

Converting conservation laws from the Lagrangian coordinates 
to the Eulerian coordinates, one should remember that 
the mass Lagrangian coordinate    $ s $ is defined 
in the Eulerian coordinates  by the equations 
\begin{equation}   \label{nonlocal} 
s_r = r \rho ,
\qquad
s_t = - r  \rho u  , 
\end{equation}
i.e. $ s $ is a  nonlocal  variable  in the Eulerian coordinates   $ ( t,r ) $.  
The system~(\ref{nonlocal}) follows from~(\ref{m_L_coordinates_1}). 
\end{remark}

\subsection{Case   $ A \neq 0$}

\subsubsection{Arbitrary  conductivity $ \sigma ( \rho , p) $}

Using direct computations, 
we find the following seven conservation laws 
for equations~(\ref{Lagrange_system})  
in the general case 
(arbitrary $ \sigma ( \rho , p)$ and arbitrary $A$):

\begin{itemize}

\item 

mass 
\begin{equation}   \label{CL_general_A_mass} 
D_t ^L 
\left(  
{1 \over \rho } 
\right) 
- 
D_{s}
\left(  
{ r u  } 
\right) 
= 0  ; 
\end{equation}

\item

momentum along $z$-axis 
\begin{equation}   \label{CL_general_A_momentum_z} 
D_t ^L 
\left(    
   w 
\right)  
-    
D_{s}
\left( 
 r H^r   H^{z}  
\right) 
= 0  ; 
\end{equation}

\item 

motion of the center of mass  along $z$-axis 
\begin{equation}    \label{CL_general_A_galileo_z} 
D_t ^L 
\left(     
 t w   - z 
\right)  
-    
D_{s}
\left( 
   t r H^r    H^{z}  
\right) 
= 0 ; 
\end{equation}

\item

angular momentum in $ (r, \theta )$-plane 
\begin{equation}   \label{CL_general_A_rotation} 
D_t ^L 
\left(    
    r v 
\right)  
-    
D_{s}
\left(  
  r ^2 H^r   H^{\theta}  
\right)  
= 0  ; 
\end{equation}

\item 

magnetic fluxes 
\begin{equation}            \label{CL_general_A_flux_theta} 
D_t ^L 
\left(   
H^{\theta}   \over  r \rho   
\right)      
- 
D_{s}
\left( 
E^z   + 
 v H^r  
\right) 
= 0  , 
\end{equation} 
\begin{equation}           \label{CL_general_A_flux_z} 
D_t ^L 
  \left(   
H^z   \over  \rho 
\right)     
+ 
D_{s}
\left(  
 r E ^{\theta} 
-  
r w H^r 
\right) 
= 0    ; 
\end{equation}

\item

energy 
\begin{multline}   \label{CL_general_A_energy} 
D_t ^L 
\left\{
{ 1   \over 2 }    (    u  ^2 + v ^2 +  w^2  ) 
 +       
  {  1 \over  \gamma  -1  } { p \over    \rho  }
+       
 {   ( H^{\theta}  ) ^2  +  ( H^{z}  ) ^2   \over  2  \rho }
\right\} 
\\
+ 
D_{s}
\left\{
      r  u  \left(  
  p   
  + 
    {  ( H^{\theta}  ) ^2  +  ( H^{z}  ) ^2   \over 2} 
 \right) 
+ r ( E ^{\theta}  H ^z  - E ^z H ^{\theta} ) 
 -  r H^r  ( v H^{\theta} + w H^{z}  ) 
\right\} 
=  0  . 
\end{multline}
\end{itemize}

\subsubsection{Special cases of  conductivity $ \sigma ( \rho , p) $}

For electric conductivity $\sigma ( \rho , p)  =  C \rho $ with constant   $ C \neq 0  $ 
there exists the additional conservation law 
\begin{equation}      \label{special_CL_1} 
D_t  ^L \left\{
    \left(2 t - C s \right) \frac{H^z}{\rho} - C r z H^r  
\right\}
+ D_s \left\{
    \left( 2 t - C s \right) (r E^{\theta} - r w H^r  )
    - r^2 H^z
\right\} = 0.
\end{equation}

\subsection{Case   $ A = 0$}

For $ A = 0$ we get $ H^r \equiv 0$.  
We will consider the reduced  system~(\ref{Lagrange_system2}).

\subsubsection{Arbitrary   conductivity $ \sigma ( \rho , p) $}

The following  conservation laws of equations~(\ref{Lagrange_system2})  
are obtained by direct computations:

\begin{itemize}

\item 

mass 
\begin{equation}   \label{CL_A0_mass} 
D_t ^L 
\left(  
{1 \over \rho } 
\right) 
- 
D_{s}
\left(  
{ r u  } 
\right)  
= 0  ; 
\end{equation}

\item

angular momentum in $ (r, \theta )$-plane 
\begin{equation}   \label{CL_A0_rotation} 
D_t ^L 
\left(    
    r v 
\right) 
= 0  ; 
\end{equation}

\item 

magnetic  fluxes 
\begin{equation}            \label{CL_A0_flux_theta}  
D_t ^L 
\left(   
H^{\theta}   \over  r \rho   
\right)     
- 
D_{s}
\left(  
E^z  
\right) 
=  0   , 
\end{equation} 
\begin{equation}          \label{CL_A0_flux_z} 
D_t ^L 
  \left(   
H^z   \over  \rho 
\right)    
+ 
D_{s}
\left(    r E ^{\theta} 
\right)  
= 0   ; 
\end{equation}

\item

energy  
\begin{multline}    \label{CL_A0_energy} 
D_t ^L 
\left\{
{ 1   \over 2 }    (    u  ^2 + v ^2  ) 
 +       
  {  1 \over  \gamma  -1  } { p \over    \rho  }
+       
 {   ( H^{\theta}  ) ^2  +  ( H^{z}  ) ^2   \over  2  \rho }
\right\} 
\\
+ 
D_{s}
\left\{
      r  u  \left(  
  p   
  + 
    {  ( H^{\theta}  ) ^2  +  ( H^{z}  ) ^2   \over 2} 
 \right) 
+ r ( E ^{\theta}  H ^z  - E ^z H ^{\theta} ) 
\right\}
=  0  .
\end{multline}
\end{itemize} 
Note that the conservation law for the angular momentum~(\ref{CL_A0_rotation})   
has trivial spacial density:   $ T^s \equiv 0$.


\begin{remark} 
If we consider the complete system which consists of~(\ref{Lagrange_system2}) 
and~(\ref{Lagrange_system2_w}), 
then we obtain more conservation laws with $ T^s \equiv 0$: 
\begin{equation}      \label{CL_A0_ODE_type} 
D_t ^L    ( T^t ( r v, w, z - t w )  ) = 0   . 
\end{equation}  
The energy conservation laws in this case should include $w$, 
i.e. it has the form  
\begin{multline}   
D_t ^L 
\left\{
{ 1   \over 2 }    (    u  ^2 + v ^2 +  w^2  ) 
 +       
  {  1 \over  \gamma  -1  } { p \over    \rho  }
+       
 {   ( H^{\theta}  ) ^2  +  ( H^{z}  ) ^2   \over  2  \rho }
\right\} 
\\
+ 
D_{s}
\left\{
      r  u  \left(  
  p   
  + 
    {  ( H^{\theta}  ) ^2  +  ( H^{z}  ) ^2   \over 2} 
 \right) 
+ r ( E ^{\theta}  H ^z  - E ^z H ^{\theta} ) 
\right\}
=  0  .
\end{multline}
\end{remark}

\subsubsection{Special cases of  conductivity $ \sigma ( \rho , p) $}

For conductivity   $\sigma ( \rho , p) =  C \rho $, $ C \neq 0 $ 
there are two additional conservation laws

\begin{equation}    \label{special_CL_2} 
D_t   ^L   \left\{
    \left( 2 t - C s \right) \frac{H^z}{\rho}
\right\}
+ D_s\left\{
    \left( 2 t - C s \right) r E^{\theta}
    - r^2 H^z
\right\}= 0
\end{equation}
and
\begin{equation}    \label{special_CL_3} 
D_t   ^L   \left(
    C  s   \frac{H^\theta}{r \rho}
\right)
- D_s\left(
    C s E^z
    - r H^\theta 
\right) = 0.
\end{equation}
Using equivalence transformations,  
one can set  $C=1$.

\begin{remark}
Conservation laws    (\ref{special_CL_1}),    (\ref{special_CL_2})   and   (\ref{special_CL_3}),  
which were obtained for special values of $  \sigma ( \rho , p) = C \rho  $, 
hold only for finite electric conductivity, 
i.e. they have no  analogs for infinite conductivity $ \sigma = \infty $. 
\end{remark}


\section{Symmetries in the case of infinite  conductivity  }

\label{Symmetries_infinite}

For infinite electric conductivity ($ \sigma = \infty$) 
the MHD equations can be derived from the system~(\ref{3D_system}).  
One needs to take the limiting case  $ \sigma  \rightarrow  \infty$.   
Taking the limiting case of equations~(\ref{Lagrange_system}),   
describing  the one-dimensional MHD  flows with cylindrical symmetry 
in the mass Lagrangian coordinates, 
we obtain  the system  
\begin{subequations}    \label{Lagrange_system3} 
\begin{gather}  
\rho_t = -\rho^2 (r u) _{{s}},    
  \label{Lagrange_system3_rho} 
\\
 u_t   -  { v^2 \over  r }  = - r p_{{s}}
-   { 1  \over 2  r  }       \left(   r^2  ( H^{\theta} )^2     \right)  _s 
-   { r  \over 2 }       \left(    ( H^z )^2     \right)  _s  ,
\qquad
r_t = u ,  
 \label{Lagrange_system3_u} 
\\
 v _t   +   {  u v  \over  r }  
=  { H^r  }     \left(   r   H^{\theta}    \right)  _s , 
\qquad 
{\theta} _t  = {  v \over r }    ,  
 \label{Lagrange_system3_v} 
\\
 w_t      
=  r { H^r }         H^{z}     _s , 
\qquad 
z_t =  w , 
 \label{Lagrange_system3_w} 
\\
   p_t  
= - \gamma  \rho p ( r u )_{{s}}  , 
  \label{Lagrange_system3_p} 
\\     
  H^{\theta}  _t  
= 
r \rho (   ( v H^r  )  _s    -  H^{\theta} u_s )  , 
 \label{Lagrange_system3_H_theta} 
\\ 
  H^z _t 
=  \rho  (     ( r  w H^r  ) _s      -   H^z  (r u )_s   )     . 
  \label{Lagrange_system3_H_z} 
\end{gather}
\end{subequations}


\begin{remark} 
Using~(\ref{Lagrange_system3_rho}), 
it is possible to rewrite equation~(\ref{Lagrange_system3_H_theta})  as 
 \begin{equation*}     
  \left(   
H^{\theta}   \over  \rho 
\right)      _t
   =  
  r  \left( v H^r \right) _s  
+ { u  H^{\theta}   \over  r  \rho   }  
 \end{equation*}
 that can be further simplified as 
\begin{equation}         \label{Modification3_H_theta}
\left(   
H^{\theta}   \over  r \rho   
\right)      _t
= 
\left( v H^r  \right) _s  . 
\end{equation}
Similarly,  use of~(\ref{Lagrange_system3_rho})   
allows to rewrite equation~(\ref{Lagrange_system3_H_z})    as 
\begin{equation}         \label{Modification3_H_z}
  \left(  
 { H^z   \over  \rho } 
\right)      _t
  =    
\left( r w H^r  \right) _s   . 
\end{equation}   


Switching to variables    $\varphi (t,s)$,     $  \psi (t,s) $  and  $    \chi (t,s)$ 
introduced by~(\ref{m_L_coordinates}),(\ref{m_L_coordinates_details}), 
we can rewrite     the conservation laws~(\ref{Modification3_H_theta}) 
and~(\ref{Modification3_H_z}) as 
\begin{equation*}         
\left(   \varphi _s  H^{\theta}     \right)      _t
= ( A \psi _t  ) _s  , 
\qquad 
  \left(  {    \varphi \varphi _s   }   H^z  \right)      _t
  =     ( A \chi _ t   ) _s   . 
\end{equation*}   
Integration leads to 
\begin{equation*}         
 \varphi _s  H^{\theta}    
=  A \psi _s  + g' _1 (s)   , 
\qquad   
  {    \varphi \varphi _s  }   H^z  
  =      A \chi _ s    + g' _2 (s)   , 
\end{equation*}   
where  $g_1 (s) $  and  $g_2 (s) $  are arbitrary functions.

If $ A\neq 0$,  then it is possible to choose functions    $\psi (t,s)$ and     $  \chi  (t,s) $ 
such  that   $g_1 (s) \equiv 0 $  and  $g_2 (s) \equiv 0 $. 
Thus we obtain 
\begin{equation}         
 H^{\theta}    
=  A { \psi _s   \over   \varphi _s }    , 
\qquad 
  H^z  
  =      A  {  \chi _ s    \over     \varphi  \varphi _s}   . 
\end{equation}   
We also get 
\begin{equation}     
 \theta _s      
=  { H^{\theta}   \over  A  r \rho   }   , 
\qquad 
  z _s  
  =     {   H^z    \over   A \rho  }   , 
\end{equation}   
or, equivalently, 
\begin{equation}        \label{define_z_s}
 \theta _s      
=  { H^{\theta}   \over  r ^2  \rho  H^r   }   , 
\qquad 
  z _s  
  =     {   H^z    \over   r  \rho  H^r  }   . 
\end{equation}   
\end{remark}

Equations~(\ref{Lagrange_system3_rho})  and~(\ref{Lagrange_system3_p}) provide
\begin{equation*}
\left(  {  p \over \rho ^{\gamma}  } \right)   _t   = 0  . 
\end{equation*}
Therefore, the entropy   function 
$
S   =  {  p \over \rho ^{\gamma}  }
$
satisfies
\begin{equation}      \label{Modification3_entropy}
S_t = 0 . 
\end{equation}   
It means that the entropy is conserved  along the trajectories of motion.

\subsection{Case   $ A \neq 0$}

The equivalence transformations of system~(\ref{Lagrange_system3})   
are provided  in~(\ref{infinite_equivalence_C_generators}),  Appendix~\ref{Equivalence}. 
Note that the system is extended by 
the nonlocal variable $r$ is defined by~(\ref{nonlocal_r}). 
The transformations  allow scaling of the constant $A$.  
The system~(\ref{Lagrange_system3}) 
admits symmetries 
\begin{multline}
X_1 = {\ddt},
\qquad
X_2 = {\dds},
\qquad 
X_3 = t  {\ddt} + 2 s {\dds}
    - u {\ddu} - v {\ddv} - w {\ddw}
    + 2 \rho {\ddrho},
\\
X_4 =
- 2 s {\dds} +  r {\ddr}  +  z {\ddz}
    +  v {\ddv} +   u {\ddu} + w {\ddw}
    -  4 \rho {\ddrho}    -  2 p {\ddp}  
 -  H^{\theta}   {\ddHtheta} -  H^z  {\ddHz} ,
\\
X_{5} = t {\ddz} + {\ddw}  , 
\qquad
X_6 =    f_1   \left(s, { p  \over \rho^{\gamma} }  \right)    {\ddtheta} , 
\qquad
X_{7} = f_2 \left(s, { p  \over \rho^{\gamma} }  \right)    {\ddz},
\end{multline}
where $ f_1 $ and $ f_2 $  are arbitrary functions.  
We remark that $ p  /   \rho^{\gamma}  $ 
can be replaced by the entropy variable $S$.


\subsection{Case   $ A = 0$}

For  $ A = 0$ the system~(\ref{Lagrange_system3})  gets simplified 
because $ H^r \equiv 0$. 
It takes the form 
\begin{subequations}    \label{Lagrange_system4} 
\begin{gather}  
\rho_t = -\rho^2 (r u) _{{s}},    
  \label{Lagrange_system4_rho} 
\\
 u_t   -  { v^2 \over  r }  = - r p_{{s}}
-   { 1  \over 2  r  }       \left(   r^2  ( H^{\theta} )^2     \right)  _s 
-   { r  \over 2 }       \left(    ( H^z )^2     \right)  _s  ,
\qquad
r_t = u ,  
 \label{Lagrange_system4_u} 
\\
 v _t   +   {  u v  \over  r }  
=  0 ,  
\qquad 
{\theta} _t  = {  v \over r }    ,  
 \label{Lagrange_system4_v} 
\\
   p_t  
= - \gamma  \rho p ( r u )_{{s}} , 
  \label{Lagrange_system4_p} 
\\     
  H^{\theta}  _t  
= 
- r \rho   H^{\theta} u_s   , 
 \label{Lagrange_system4_H_theta} 
\\ 
  H^z _t 
=  -  \rho    H^z  (r u )_s     .  
  \label{Lagrange_system4_H_z} 
\end{gather}
\end{subequations}
In this case the equations 
\begin{equation}
w_t      
=  0 , 
\qquad 
z_t =  w , 
 \label{Lagrange_system4_w} 
\end{equation}
can be separated as it was done in the case of finite conductivity 
for system~(\ref{Lagrange_system2}),(\ref{Lagrange_system2_w}).   
These equations were solved in~(\ref{w_solved}).

As remarked in point~\ref{finite_sigma_A_0} 
there is a possibility for further  simplification  of the system 
if  we assume that the motion is radial, i.e. $ v (t,s) \equiv  0$.

\begin{remark} 
Using~(\ref{Lagrange_system4_rho}), 
it is possible to rewrite equation~(\ref{Lagrange_system4_H_theta})  as 
 \begin{equation*}     
  \left(   
{ H^{\theta}   \over  \rho }
\right)      _t
   =  
{ u  H^{\theta}   \over  r  \rho   }  
 \end{equation*}
 that can be further simplified as 
\begin{equation}         \label{Modification4_H_theta}
\left(   
{ H^{\theta}   \over  r \rho  }  
\right)      _t
= 0   . 
\end{equation}
Also using~(\ref{Lagrange_system4_rho}), 
we rewrite equation~(\ref{Lagrange_system4_H_z})    as 
\begin{equation}         \label{Modification4_H_z}
  \left(   
{ H^z   \over  \rho } 
\right)      _t
  =    0   . 
\end{equation}   
\end{remark}


The  system~(\ref{Lagrange_system4})  admits symmetries 
\begin{multline}
X_1 = {\ddt},
\qquad
X_2 = {\dds},
\qquad 
X_3 = t  {\ddt} + 2 s {\dds}
    - u {\ddu} - v {\ddv} 
    + 2 \rho {\ddrho},
\\
X_4 =
- 2  s {\dds} +  r {\ddr}  
    +  v {\ddv} +   u {\ddu} 
    -  4 \rho {\ddrho}  -  2 p {\ddp}  
-  H^{\theta}   {\ddHtheta} -  H^z  {\ddHz} ,  
\\
X_5 =
  2  s {\dds} 
    + 2  \rho {\ddrho}  +   2 p {\ddp} 
  +  H^{\theta}   {\ddHtheta}  +   H^z  {\ddHz}  , 
\qquad
X_6 =    f _1    {\ddtheta} ,  
\end{multline}
where 
\begin{equation*} 
f_1 =  f _1 \left(   s, r v,   { p \over \rho ^{ \gamma}  } ,   { H^{\theta}  \over r \rho}   ,  { H^{z}  \over \rho}  \right) 
\end{equation*}
is an arbitrary function.

For $\gamma=2$ there is an extension by the generator
\begin{equation} 
X_7 
= 
 \rho    f_{2}  
 \left(   {\ddHz}    -    H^{z}    {\ddp}   \right) 
\end{equation} 
with  arbitrary function 
\begin{equation*} 
f  _{2} = f _2 \left(   s, r v,   { p \over \rho ^{ \gamma}  } ,   
{ H^{\theta}  \over r \rho}   ,  { H^{z}  \over \rho}  \right)  .
\end{equation*}


\section{Conservation laws  in the case of infinite conductivity: variational approach}

\label{Conservation_infinite}

In the case of infinite electric  conductivity we can bring 
the equations for one-dimensional MHD flows 
with cylindrical symmetry 
into a variational form.  
This possibility can be exploited to find conservation laws: 
one can employ the Noether theorem.  
To the best of our knowledge it is not possible 
for the case of finite electric conductivity.

The decisive property of the infinite conductivity case, 
which allows to bring  the equations into a variational form, 
is the  conservation of entropy~(\ref{Modification3_entropy}). 
We remark that in the case of finite conductivity 
we have (in the mass  Lagrangian coordinates)
\begin{equation*}  
 S _ t    
    =
      {    \gamma  -1    \over \rho   ^{\gamma} }    \   (  \textbf{i}     \cdot   \textbf{E}  )   .
\end{equation*}


\subsection{Case $ H^r  = A / r $ with $ A \neq 0$}

\subsubsection{Variational  formulation}

First we consider the case $ A \neq 0$ that implies $ H^r   {\not \equiv}   \    0  $. 
The system~(\ref{Lagrange_system3}), 
which describes one-dimensional MHD flows 
with cylindrical symmetry,  takes the form 
\begin{subequations}    \label{Variational_system}
\begin{gather} 
\left(  {  1 \over  \rho  }  \right) _t =  (r u) _{{s}},    
 \label{Variational_system_rho}
\\
 u_t   -  { v^2 \over  r }  = - r p_{{s}}
-   { 1  \over 2 r  }       \left(   r^2  ( H^{\theta} )^2     \right)  _s 
-   { r  \over  2   }       \left(    ( H^z )^2     \right)  _s  ,
\qquad
r_t = u ,  
 \label{Variational_system_u}
\\
 v _t   +   {  u v  \over  r }  
=  { H^r }     \left(   r   H^{\theta}    \right)  _s , 
\qquad 
\theta _t =  {  v  \over  r }      , 
 \label{Variational_system_v}
\\
 w_t      
=  r { H^r }         H^{z}   _s , 
\qquad 
z_t =  w , 
 \label{Variational_system_w}
\\
S_t = 0  , 
 \label{Variational_system_S}
\\
 \left(  H^{\theta}   \over  r \rho   \right)      _t
 = ( v H^r  )_s  ,
 \label{Variational_system_H_theta}
\\     
  \left(   H^z   \over  \rho \right)      _t
  =     ( r w H^r  ) _s   .
 \label{Variational_system_H_z}
\end{gather}
\end{subequations}
Here we used equation~(\ref{Modification3_entropy}) for the entropy 
instead of the equation for the pressure 
and  equations~(\ref{Modification3_H_theta})  and~(\ref{Modification3_H_z}) 
instead of equations for components of the magnetic  field  $ H^{\theta}$  and $ H^z$.


Employing functions $ \varphi  (t,s) $,  $ \psi (t,s) $ and  $ \chi  (t,s) $,  
described in~(\ref{m_L_coordinates}),(\ref{m_L_coordinates_details}), 
we express the physical variables as
\begin{subequations}    \label{subs_general_A_0} 
\begin{gather} 
\label{subs_general_A_1} 
u = \varphi   _ t   ,
\qquad
  \rho =     {  1  \over  \varphi    \varphi  _s  }    , 
\\
      \label{subs_general_A_2} 
v     =    {   \varphi   \psi  _t    }  , 
\qquad 
H^{\theta} 
 =  A  {   \psi  _s    \over \varphi _s }   , 
\\
   \label{subs_general_A_3} 
 w     =    {  \chi  _t     }  , 
\qquad
H^{z} 
 =     A  { \chi   _s    \over     \varphi     \varphi _s }    .
\end{gather}
\end{subequations}
These substitutions  for the physical variables make 
equations~(\ref{Variational_system_rho}),   (\ref{Variational_system_H_theta}) 
and~(\ref{Variational_system_H_z}) satisfied.   
The entropy equation~(\ref{Variational_system_S}) 
can be solved as 
\begin{equation}     \label{subs_general_A_4} 
S= S(s)  , 
\end{equation}
where  $S(s)$ is an arbitrary function.

We are left  with equations~(\ref{Variational_system_u}),  
(\ref{Variational_system_v}) 
and~(\ref{Variational_system_w}). 
They  get rewritten as three second-order PDEs
\begin{subequations}     \label{variational_three_PDEs}
\begin{gather} 
\varphi  _{t t }
-   \varphi          \psi  _t    ^2 
=
 -    \varphi
  \left(   {  S  \over     \varphi   ^{\gamma}  \varphi  _{s}  ^{\gamma}  }       \right)    _{{s}}
- {  A ^2      \over 2  \varphi     }
  \left(    { \varphi  ^2       \psi  _s  ^2    \over \varphi _s  ^2  }   \right)   _{{s}}   
- {  A ^2   \varphi       \over  2    }
  \left(    {     \chi   _s  ^2    \over   \varphi  ^2   \varphi _s  ^2  }   \right)   _{{s}}     , 
\\
( \varphi      \psi  _{t} )    _t   
+   {   \varphi   _t     \psi  _t   } 
= 
 {  A^2     \over  \varphi   }
  \left(    { \varphi        \psi  _s    \over \varphi _s   }   \right)   _{{s}}    , 
\\
  \chi   _{t t} 
= 
 {  A^2        }
  \left(    {       \chi  _s    \over    \varphi  \varphi _s   }   \right)   _{{s}}    .
\end{gather}
\end{subequations}
This system of PDEs is variational. 
It is provided by the Lagrangian function 
\begin{equation}   \label{Lagrangian_1}
L
=    { 1\over 2 }  
\left( 
  \varphi  _{t}   ^2
+     \varphi     ^2     \psi  _t    ^2 
+        \chi  _t   ^2  
\right) 
 -   {  S  \over      \gamma  -1 }     \varphi    ^{1 -  \gamma}       \varphi  _{s}  ^{1 -  \gamma}
- {  A^2   \over 2 } 
 \left( 
   {  \varphi      \psi  _s  ^2 \over  \varphi  _s  }   
+  {    \chi   _s  ^2 \over  \varphi \varphi  _s  }  
\right)    , 
\end{equation}
i.e. equations~(\ref{variational_three_PDEs})  
are Euler--Lagrange equations  for this Lagrangian.  

\begin{remark}
The Lagrangian function~(\ref{Lagrangian_1}) has a standard physical interpretation.  
If rewritten in the physical variables, 
it gets presented as the kinetic energy  minus the potential energy: 
\begin{equation*}
L
=  { 1   \over 2 }    (    u  ^2 + v ^2 +  w^2  ) 
 -        {  S \over  \gamma  -1  }    \rho  ^{ \gamma -1 }
-          {   ( H^{\theta}  ) ^2  +  ( H^{z}  ) ^2   \over  2  \rho } . 
\end{equation*}
The  potential energy consists of the internal  energy of the medium and the magnetic energy. 
\end{remark}

\begin{remark}
The Lagrangian~(\ref{Lagrangian_1}) does not have  the energy  term corresponding to $H^r$. 
This term, namely  
\begin{equation*}
- 
    {   ( H^{r}  )  ^2   \over  2  \rho } 
= 
- {  A^2          \varphi  _{s}  \over   2    \varphi     }   , 
\end{equation*}
is a total divergence. 
It  does  not contribute to the Euler--Lagrange  equations~(\ref{variational_three_PDEs}): 
total divergences are annihilated by variational operators. 
\end{remark}

\subsubsection{Symmetries}

\label{Symmetries_A_not_zero}

The equivalence transformations  of system~(\ref{variational_three_PDEs}) 
are provided in~(\ref{equivalence_S}), Appendix~\ref{Equivalence}.
They allow to scale the entropy function $ S(s) $ 
and constant $A$.

Symmetries of~(\ref{variational_three_PDEs})  
can be  presented as 
\begin{equation}      \label{generator_form_1}
X =\sum _{i=1} ^8   k_i Y_i   , 
\end{equation}
where 
\begin{multline} 
Y_1 =  {\ddt} , 
\qquad 
Y_2 =  {\dds} , 
\qquad 
Y_3 =  { \partial  \over \partial \psi   } , 
\qquad 
Y_4 = { \partial  \over \partial \chi  }  , 
\qquad
Y_5 =  t { \partial  \over \partial \chi  }   , 
\\
Y_6 =   t {\ddt}  , 
\qquad 
Y_7 =  s {\dds}  , 
\qquad 
Y_8 =   \varphi  {\ddphi} +  \chi  { \partial  \over \partial \chi  }   .
\end{multline}

Application of operator~(\ref{generator_form_1}) 
to  system~(\ref{variational_three_PDEs}) 
provides the conditions on the coefficients  $ k_1,  \ldots , k_8 $:    
\begin{subequations}     \label{conditions_1}
\begin{gather}
( k_7 s + k_2 )  S_s  (s)
= 
 ( -2 k_6 + ( 1 - \gamma) k_7 + 2 \gamma  k_8) S  (s)    ,
\label{conditions_1a}
\\
2 k_6  =  k _7 + 2 k_8     .
\label{conditions_1b}
\end{gather}
\end{subequations}
The first condition stands as a classifying equation for  $ S (s) $. 
We can rewrite it as 
\begin{equation}      \label{classifying_form}
(  \alpha _1 s  +  \alpha _0)   S_s  (s) =   \beta S   (s) 
\end{equation}
with constant parameters   $ \alpha _0 $, $ \alpha _1 $  and $ \beta $.  
This equation was also obtained in analysis of gas dynamics  
equations~\cite{bk:AndrKapPukhRod[1998], DORODNITSYN2019201, DorodnitsynKozlovMeleshko2011}  
and for plain one-dimensional MHD flows~\cite{DKKMM2021}.   
The equation leads to four cases of the entropy function $ S (s) $. 
They are 
\begin{itemize}

\item

arbitrary   $ S(s) $;

\item

constant $ S(s) =S_{0} $,  $ S_{0} = \mbox{const}$;

\item

power $ S (s) =S_{0} s^{q}  $,  $ q \neq 0 $, $ S_0   = \mbox{const} $;

\item

exponential $  S = S_{0} e^{q s}  $,  $  q\neq 0 $, $ S_0   = \mbox{const}    $.

\end{itemize}
Thus, we have to consider the general case with arbitrary $ S (s) $ 
and three special cases.


From~(\ref{conditions_1}) we get {four} symmetries for arbitrary $ S (s)$:  
\begin{equation}       \label{four_symmetries}
X_1 = {\ddt} , 
\qquad 
X_2  = { \partial  \over  \partial \psi} , 
\qquad 
X_3  = { \partial  \over  \partial \chi } , 
\qquad 
X_4  = t { \partial  \over  \partial \chi }  . 
\end{equation}
It is the kernel of the Lie algebras admitted by the system.     
We remark that these operators form a subset of operators~(\ref{kern01}).


Additional symmetries 
which are admitted for particular cases 
of the  entropy function $ S(s)$ are presented  in  Table 3.

\begin{equation*}
\begin{array}{|c|c|l|l|}
\hline
\mbox{Case} &    S (s)
& \mbox{Symmetry Extension}
& \mbox{Conditions}
 \\
\hline
 & & & \\
1    &   S_0    &
{   \displaystyle
{\dds} ,
\qquad
( 2  \gamma - 1 ) t {\ddt} 
+ 2( \gamma -1) s {\dds}
+ \gamma \left( \varphi  {\ddphi}
+  \chi  { \partial  \over \partial \chi  }  \right)    }  &
\\
 & & & \\
\hline
 & & & \\
2       &   S_0  s^q    &
{   \displaystyle 
 (2\gamma+q - 1 )  t {\ddt}
+ 2 (\gamma - 1 ) s {\dds}
+  (\gamma + q )
\left(  \varphi  {\ddphi}
+  \chi  { \partial  \over \partial \chi  }
\right)   }
 &
q \neq 0
\\
 & & & \\
\hline
 & & &  \\
3    &   S_0  e^{q s}   &
{   \displaystyle
 q   t {\ddt}
+ 2 ( \gamma  -1) {\dds}
+ q
\left(  \varphi  {\ddphi}
+  \chi  { \partial  \over \partial \chi  }
\right)   }
 &
q \neq 0
\\
 & & & \\
\hline
\end{array}
\end{equation*}
\begin{center}
{Table 3:} Additional  symmetries  for $ A \neq 0$.
$ S_0 $ is a nonzero constant. 
\end{center}


\subsubsection{Conservation laws}

\noindent 
{\bf a) Arbitrary  $ S(s)$}

For arbitrary entropy function $ S(s) $  Lagrangian~(\ref{Lagrangian_1}) 
possesses four symmetries~(\ref{four_symmetries}). 
Symmetries $ X_1 $,    $ X_2 $ and    $ X_3 $  are variational. 
Symmetry $ X_4 $ is a divergence symmetry with  $ ( B_1 , B_2 ) =  (\chi , 0) $. 
These symmetries provide  conservation laws of 
energy, 
angular momentum, 
momentum along  $z$-axis 
and 
motion of the   center of mass along $z$-axis. 
These conservation laws also exist in the case of finite electric conductivity, 
for which they are given 
by~(\ref{CL_general_A_energy}), 
(\ref{CL_general_A_rotation}), 
(\ref{CL_general_A_momentum_z})
and~(\ref{CL_general_A_galileo_z}), 
respectively.

The conservation laws which were used to bring  the system 
into the  variational form, 
i.e.  conservation of mass, 
conservation laws for  
magnetic fluxes   
and conservation of entropy, 
cannot be obtained from the Lagrangian.     
First three of these conservation  laws  also   exist in the case of finite  electric conductivity. 
There are given by~(\ref{CL_general_A_mass}), 
(\ref{CL_general_A_flux_theta})    
and~(\ref{CL_general_A_flux_z}), 
respectively. 
The conservation of entropy~(\ref{Modification3_entropy}) 
holds only in the case of infinite  electric conductivity.

We remark that we obtained the same conservation laws 
as  we found in the case of finite electric conductivity 
and   conservation law for the entropy.

\medskip

\noindent 
{\bf b) Special cases of $ S(s)$  }

The additional variational symmetries  of the Lagrangian, 
which exits for special cases of the entropy function  $S(s)$,  
are given in Table~4.

\begin{equation*}
\begin{array}{|c|c|l|l|}
\hline
\mbox{Case} &    S(s)
& \mbox{Symmetry Extension}
& \mbox{Conditions}
 \\
\hline
 & & & \\
1    &   S_0     &
{   \displaystyle
{\dds}   }  &
\\
 & & & \\
\hline
 & & & \\
2    &   S_0  s^q    &
{   \displaystyle
 2s {\dds} 
-  \left(  \varphi  {\ddphi}
+    \chi  { \partial  \over \partial \chi  }   \right)   }
 &
{   \displaystyle
q =1 - 2  \gamma  }
\\
 & & & \\
\hline
\end{array}
\end{equation*}
\begin{center}
{Table 4:}    Additional variational symmetries   for  $ A \neq 0 $.
$ S_0 $ is a nonzero constant. 
\end{center}


There are two particular cases $ S(s)$ for which there exist  
additional symmetries of the Lagrangian.  
Using these symmetries, 
we obtain the following conservation laws.

\begin{itemize} 

\item 

Case  $ S  (s) = S_0 $.  

In the case of constant  entropy $ S  (s) = S_0 $  
there exists the additional variational symmetry 
\begin{equation*}
 { \partial \over \partial s  }   . 
\end{equation*}
It provides the conservation law 
\begin{equation*}     
D_t ^L 
\left(     
     \varphi  _{t}       \varphi  _{s}   
+
    \varphi   ^2    \psi  _{t}    \psi  _{s}  
+ 
    \chi   _t      \chi   _s  
\right)  
+     
D_s
\left( 
 -  { 1\over 2 }  
\left( 
  \varphi  _{t}   ^2
+     \varphi     ^2     \psi  _t    ^2 
+        \chi  _t   ^2  
\right) 
 +    {  \gamma S  \over      \gamma  -1 }     \varphi    ^{1 -  \gamma}       \varphi  _{s}  ^{1 -  \gamma}
\right)  
= 0   .
\end{equation*}
We can rewrite it in the physical variables as 
\begin{equation}     \label{conservation_general_A_s_physical} 
D_t ^L 
\left(     
  { u   H^r      +     v  H^{\theta}    +     w H^z    \over  r \rho  H^r }   
\right)  
+     
D_s
\left( 
 - 
{ 1   \over 2 }    (    u  ^2 + v ^2 +  w^2  ) 
 +      
  {  \gamma S \over  \gamma  -1  }    \rho  ^{ \gamma -1 }
\right)  
= 0   . 
\end{equation}


\item

Case  $ S  (s) = S_0  s^q $ with $  q = 1 - 2 \gamma$.

In the case of power entropy with  $  q = 1 - 2 \gamma$, 
there exists the additional variational symmetry 
\begin{equation*} 
  2 s  { \partial \over \partial s  }   
- \left( \varphi  { \partial \over \partial   \varphi  }  
+  \chi  { \partial \over \partial   \chi  }  \right) , 
\end{equation*}
which leads to  conservation law 
\begin{multline*}
D_t ^L 
\left\{ 
 2 s 
(   \varphi  _{t}   \varphi _s    
+    \varphi ^2    \psi  _{t}    \psi   _s    
+   \chi  _{t}    \chi _s     )  
+  
\varphi    \varphi  _{t} 
+ 
\chi   \chi  _{t}   
\right\}
\\
+ 
D_s 
\left\{ 
2s 
\left(
-    { 1\over 2 }  
\left( 
  \varphi  _{t}   ^2
+     \varphi     ^2     \psi  _t    ^2 
+        \chi  _t   ^2  
\right) 
 +    {    \gamma  S  \over      \gamma  -1 }     \varphi    ^{1 -  \gamma}       \varphi  _{s}  ^{1 -  \gamma}
\right)
\right.
\\
\left. 
+ 
\varphi  
\left[ 
  {  S    }     \varphi    ^{1 -  \gamma}       \varphi  _{s}  ^{ -  \gamma}
 +  {  A^2   \over 2 } 
 \left( 
   {  \varphi      \psi  _s  ^2 \over  \varphi  _s  ^2 }   
+  {    \chi   _s  ^2 \over  \varphi \varphi  _s ^2  }  
\right)  
\right]
-  
{  A^2  } 
  {   \chi   \chi   _s   \over  \varphi \varphi  _s  }  
\right\} = 0  .
\end{multline*}
It takes the form
\begin{multline}      \label{CL_scaling_with _A} 
D_t ^L 
\left\{ 
2 s 
{ u   H^r      +     v  H^{\theta}    +     w H^z    \over  r \rho  H^r }   
+  
r u 
+ 
z w   
\right\}
\\
+ 
D_s 
\left\{ 
2s 
\left(
 -  { 1   \over 2 }    (    u  ^2 + v ^2 +  w^2  ) 
 +         { \gamma  S \over  \gamma  -1  }    \rho  ^{ \gamma -1 }
\right)
\right.
\\
\left. 
+ 
r ^2 
\left(
  {  S    }    \rho  ^{\gamma}     
+  
{    (  H ^{\theta} )^2    +    (  H^z ) ^2 \over 2 } 
\right)
-  
 r 
{  H^r  } 
z
  H ^z  
\right\} = 0
\end{multline}
in  the physical    variables.




\end{itemize}

It should be noted that
for these conservation laws  
we also need the equations~(\ref{nonlocal_r}), 
which define $r$ as a nonlocal variable in the Lagrangian coordinate system $ (t.s)$.  
For conservation law~(\ref{CL_scaling_with _A}) 
we also need  nonlocal variable $z$ which 
is defied by  the second equations in~(\ref{Variational_system_w}) 
and~(\ref{define_z_s}).

\subsection{Case $  H^r  \equiv 0 $ ($ A=  0$)}

\subsubsection{Variational formulation}

For $  A = 0 $  we get $ H ^r  \equiv 0 $. 
The system~(\ref{Lagrange_system4}) 
can be rewritten as  
\begin{subequations}    \label{Variational_system2}
\begin{gather} 
\left( { 1  \over \rho  } \right) _t  =   (r u) _{{s}},    
 \label{Variational_system2_rho}
\\
 u_t   -  { v^2 \over  r }  = - r p_{{s}}
-   { 1  \over 2 r  }       \left(   r^2  ( H^{\theta} )^2     \right)  _s 
-   { r  \over  2   }       \left(    ( H^z )^2     \right)  _s  ,
\qquad
r_t = u ,  
 \label{Variational_system2_u}
\\
 ( r v )  _t   =  0 , 
\qquad 
\theta _t =  {  v  \over  r }      , 
 \label{Variational_system2_v}
\\
S_t = 0  , 
 \label{Variational_system2_S}
\\
 \left(  { H^{\theta}   \over  r \rho   } \right)      _t
 = 0  ,
 \label{Variational_system2_H_theta}
\\     
  \left(   {  H^z   \over  \rho }  \right)      _t
  =  0    .
 \label{Variational_system2_H_z}
\end{gather}
\end{subequations}
Here we took equation~(\ref{Modification3_entropy})  
for the entropy 
instead of the equation for the pressure and 
equations~(\ref{Modification4_H_theta})  and~(\ref{Modification4_H_z})   
for the magnetic field.  
Equations~(\ref{Lagrange_system4_v})   are 
rewritten as~(\ref{Variational_system2_v}).  
We recall that the Eulerian coordinate $r$ is defined 
in  the  mass Lagrangian coordinates $ ( t,s) $ 
by the system~(\ref{nonlocal_r}).


\begin{remark} 
The system~(\ref{Variational_system2}) has conservation laws with $ T^s \equiv  0 $: 
\begin{equation}        \label{CL_special_2a}
D _t  ^L   \left\{  
T^t   \left(  r v, S,  { H^{\theta}   \over  r \rho   }  ,   {  H^z   \over  \rho }  \right) 
\right\}  = 0  . 
\end{equation}  
If we consider the system~(\ref{Variational_system2}) 
extended by equations~(\ref{Lagrange_system4_w}), 
we get more  general conservation laws of this type: 
\begin{equation*}     
D _t  ^L   \left\{  
 T^t   \left(  r v,     S,  { H^{\theta}   \over  r \rho   }  ,   {  H^z   \over  \rho }  ,  w, z - t w  \right) 
\right\} = 0  .  
\end{equation*}  
\end{remark}

Switching to function   $ \varphi ( t,s) \equiv r  $ defined in~(\ref{m_L_coordinates_1}), 
we can reduce system~(\ref{Variational_system2}) to one second-order PDE 
for the  dependent variable  $ \varphi ( t,s) $.    
Use of  
\begin{equation}     \label{subs_A0_1}
u = \varphi   _ t   ,
\qquad
  \rho =     {  1  \over  \varphi    \varphi  _s  }    
\end{equation}
makes equation~(\ref{Variational_system2_rho}) satisfied.  
The first equation in~(\ref{Variational_system2_v})  
provides 
\begin{equation}     \label{subs_A0_5}
v  = { R (s)  \over  \varphi  }  . 
\end{equation}
Equation~(\ref{Variational_system2_S}) is solved as 
\begin{equation}     \label{subs_A0_2}
S = S(s)     . 
\end{equation} 
Equations~(\ref{Variational_system2_H_theta}) 
and~(\ref{Variational_system2_H_z})   give  
\begin{equation}     \label{subs_A0_3}
H^{\theta}  = r \rho F (s)  =  {  F (s)    \over    \varphi  _s  }   , 
\qquad 
H^z  =   \rho G(s)    =      {  G(s)    \over  \varphi    \varphi  _s  }    . 
\end{equation}
Here $ R(s) $,   $ S(s) $,  $ F(s) $ and  $ G(s) $ are arbitrary smooth functions 
such that 
\begin{subequations}    \label{conditions_S_F_G} 
\begin{gather}
S (s)  \neq 0      \label{condition_S} , 
\\
F^2 (s) + G^2 (s)   {\not \equiv}  0  .      \label{conditions_F_G} 
\end{gather}
\end{subequations}
These conditions require presence of the medium and presence of the magnetic field, 
respectively.


Now the remaining equation~(\ref{Variational_system2_u})    can be rewritten as 
\begin{equation}   \label{variational_PDE}
\varphi  _{t t }  -   { R ^2  \over \varphi ^3} 
=
 -    \varphi
  \left(   {  S  \over     \varphi   ^{\gamma}  \varphi  _{s}  ^{\gamma}  }       \right)    _{{s}}
- {  1      \over 2   \varphi     }
  \left( {  F   ^2   \varphi   ^2   \over  \varphi  _s ^2 }     \right)    _{{s}}
- {  \varphi       \over 2  }
  \left( {  G  ^2   \over  \varphi   ^2   \varphi  _s ^2 }        \right)    _{{s}}   .
\end{equation}
This  second-order PDE is the Euler--Lagrange equation 
for the  Lagrangian function
\begin{equation}    \label{variational_L}
L
=    { 1\over 2 }     \varphi  _{t}   ^2   
-      { 1\over 2 }    { R ^2  \over    \varphi ^2 }
 -    {  S   \over      \gamma  -1 }     \varphi    ^{1 -  \gamma}       \varphi  _{s}  ^{1 -  \gamma}
-     {  F   ^2   \varphi    \over   2   \varphi  _s }  
-   {  G   ^2   \over   2      \varphi        \varphi  _s  }     .
\end{equation}


In the case  $\gamma =2$ both  the PDE and the Lagrangian get simplified as  
\begin{equation}   \label{variational_PDE_gamma}
\varphi  _{t t }  -   { R ^2  \over \varphi ^3} 
=
 -    \varphi
  \left(   {  \tilde{S}   \over     \varphi   ^{2}  \varphi  _{s}  ^{2}  }       \right)    _{{s}}
- {  1      \over 2   \varphi     }
  \left( {  F   ^2   \varphi   ^2   \over  \varphi  _s ^2 }     \right)    _{{s}}   
\end{equation}
and 
\begin{equation}   \label{variational_L_gamma}
L
=    { 1\over 2 }     \varphi  _{t}   ^2   
-      { 1\over 2 }    { R ^2  \over    \varphi ^2 }
 -    {  \tilde{S}  \over     \varphi          \varphi  _{s}   }
-     {  F   ^2   \varphi    \over   2   \varphi  _s }      , 
\end{equation}
where 
\begin{equation}     \label{define_tilde_S} 
\tilde{S} 
= {   S \over \gamma -1}    +  { G ^2  \over 2 }  . 
\end{equation}
For the cases  $\gamma \neq 2$  and  $\gamma = 2$ 
there are different invariance and conservation properties. 
For this reason these cases will be examined separately.

\begin{remark} 
Is physical variables Lagrangians~(\ref{variational_L}) 
and~(\ref{variational_L_gamma})
 have the form 
\begin{equation*}
L
=    {    1  \over 2 }     (  u  ^2 - v ^2   ) 
 -        {   S    \over    \gamma  -1  }  \rho  ^{\gamma -1} 
-          {  ( H^{\theta}  ) ^2   +   ( H^z ) ^2   \over  2   \rho }  . 
\end{equation*}
Note that we consider one evolutionary PDE for the dependent variable  $ \varphi $. 
Therefore,  the term  $ v ^2 /2 $  contributes to the potential energy 
in  the Lagrangian function. 
\end{remark}

\subsubsection{Equivalence transformations}

Equivalence transformations of PDEs~(\ref{variational_PDE})  
and~(\ref{variational_PDE_gamma})  
are provided Appendix~\ref{Equivalence}.  
The equivalence transformations~(\ref{equivalence_general}) allows to scale any 
three of the four functions $ \{ S, F, G, R \} $ in the PDE~(\ref{variational_PDE})  
for $ \gamma \neq 3 $. 
For  $ \gamma =  3 $ one can scale function  $G$  and any two of the functions $ \{ S, F, R \} $.  
The equivalence transformations~(\ref{equivalence_gamma_2})  can be used to scale 
all three functions $ \{ \tilde{S}, F, R \} $ of the PDE~(\ref{variational_PDE_gamma}).

\subsubsection{Symmetries in the case  $\gamma \neq 2$ ($\gamma >1$) }

Symmetries of PDE~(\ref{variational_PDE}) 
can be given as 
\begin{equation}       \label{generator_form_2}
X =\sum  _{i=1} ^5  k_i Y_i   , 
\end{equation}
where 
\begin{equation}   
Y_1 =  {\ddt} , 
\qquad 
Y_2 =  {\dds} , 
\qquad 
Y_3 =   t {\ddt}  , 
\qquad 
Y_4 =  s {\dds}  , 
\qquad 
Y_5 =   \varphi  {\ddphi}  . 
\end{equation}

Application of  operator~(\ref{generator_form_2}) 
to PDE~(\ref{variational_PDE})  provides the conditions
\begin{subequations}     \label{conditions_2}
\begin{gather}
( k_4 s + k_2 )  S_s   (s)
=  ( -2 k_3 + ( 1 - \gamma ) k_4 + 2 \gamma k_5 )  S  (s)     ,
\label{conditions_2a}
\\
( k_4 s + k_2 )  F_s (s)
=  \left( - k_3 - { 1\over 2 }   k_4 +  k_5 \right)  F (s)  , 
\label{conditions_2b} 
\\
( k_4 s + k_2 )  G_s (s)
=  \left( - k_3 - { 1\over 2 }   k_4 +  2 k_5 \right)  G (s) , 
\label{conditions_2c}
\\
( k_4 s + k_2 )  R_s (s)
=  \left( - k_3  +  2 k_5 \right)  R (s) , 
\label{conditions_2d}
\end{gather}
\end{subequations}
for coefficients $ k_1, \ldots, k_5$.

Conditions~(\ref{conditions_2}) for functions 
$ S(s) $, $ F(s) $, $ G(s) $ and $ R(s) $ have the same form 
as the classifying  equations~(\ref{classifying_form}). 
If we consider such equations for $ S $, $ F $, $ G $ and $ R $ independently, 
we obtain the same four cases 
(arbitrary, constant, power and exponential) for each of these functions
as we obtained for function  $ S(s) $ in point~\ref{Symmetries_A_not_zero}.   
However, not all possible combinations  of these cases lead to additional symmetries. 
We get only the following four combinations for consideration of the admitted symmetries:  
\begin{enumerate}

\item

Arbitrary  functions 
\begin{equation}      \label{case_arbitrary} 
S(s) , 
\quad 
F(s) , 
\quad 
G(s) , 
\quad 
R(s) ;
\end{equation} 

\item

Constant functions  
\begin{equation}    \label{case_constant} 
S(s) =S_{0} , 
\quad 
F(s) =F_{0} , 
\quad 
G(s) =G_{0} , 
\quad 
R(s) =R_{0}  ;
\end{equation}

\item

Power functions  
\begin{equation}      \label{case_power} 
S (s) =S_{0} s^{q_1}  , 
\quad 
F (s) = F_{0} s^{q_2}  , 
\quad 
G(s) =G_{0} s^{q_3}  , 
\quad  
R (s) =R_{0} s^{q_4} ; 
\end{equation}

\item

Exponential functions 
\begin{equation}      \label{case_exponential} 
S (s) =S_{0} e^{q_1 s}  ,  
\quad  
F (s) = F_{0} e^{q_2 s} , 
\quad
G(s) =G_{0} e^{q_3 s}  , 
\quad  
R (s) =R_{0} e^{q_4 s}  .
\end{equation}

\end{enumerate}
Here $ S_{0} $, $ F_{0} $,  $ G_{0} $ and $ R_{0} $  are constants. 
In all these cases we assume that  
the requirements~(\ref{conditions_S_F_G}) are satisfied. 
For  cases with constant, power and exponential functions 
it means that  
\begin{subequations}    \label{conditions_S_F_G_special} 
\begin{gather}
S _0  \neq 0    ,   \label{condition_S_special} 
\\
F _0 ^2 + G_0^2  \neq    0  .      \label{conditions_F_G_special} 
\end{gather}
\end{subequations}
For the last two cases we also assume   
\begin{equation}    \label{condition_q} 
q_1 ^2 + q_2 ^2 +  q_3 ^2 + q_4 ^2    \neq 0 . 
\end{equation}

\medskip

\noindent
{\bf  a)  Arbitrary functions case~(\ref{case_arbitrary})}

In the most general case with arbitrary functions   
$S(s)$, $F(s)$, $G(s) $ and $R(s)$  
there is only one symmetry
\begin{equation}      \label{translation} 
X_1 = {\ddt}   . 
\end{equation}

\medskip

\noindent
{\bf Special cases of  $ S(s)  $,       $ F(s) $,     $ G(s) $  and    $ R(s)  $  }

In the special cases only symmetries which exist in addition to symmetry of the case with 
arbitrary  $ S(s)  $,       $ F(s) $,     $ G(s) $  and    $ R(s)  $ are given.

\medskip

\noindent
{\bf b)   Constant functions case~(\ref{case_constant})}


For all constants 
$S_0$, $ F_0$, $ G _0$ and $R _0$ 
there exists the symmetry
\begin{equation}
X_{2}  
=
{\dds}  .
\end{equation}

For particular values of the constants there can be additional symmetries:

\begin{itemize} 

\item 

Case  
$ F_0 \neq 0 $,   $ G_0 = 0 $,  $ R_0  = 0 $   

There is one more additional symmetry 
\begin{equation}    \label{symmetry_case_2_2} 
X_{3} 
= -  t  {\ddt} 
+ 2 ( \gamma -1 ) s {\dds} 
+  ( \gamma  - 2  )  \varphi  {\ddphi}   . 
\end{equation}

\begin{itemize} 

\item

{Sub}case   $\gamma = { 3 \over 2 } $

 For $\gamma = { 3 \over 2 } $ 
this additional symmetry is  also admitted  for  $ R_0 \neq 0 $.

\end{itemize}

\item  

Case  $ F_0  = 0 $,  $ G_0 \neq 0 $, $ R_0  = 0 $ 

There is an additional  symmetry 
\begin{equation}     \label{symmetry_case_2_3} 
X_{3}  
= t  {\ddt} 
+   2 s {\dds} 
+ \varphi  {\ddphi}    .
\end{equation}

\end{itemize}

\medskip

\noindent
{\bf  c)  Power functions case~(\ref{case_power})  }

\begin{itemize}

\item  

Case 
$ F_0 \neq 0 $    (any $G_0$   and any $R_0$)

There is the additional symmetry 
\begin{equation}      \label{symmetry_case_3_1} 
X_{2}  
= ( q_1 -   2  \gamma q_2 - 1 )    t  {\ddt} 
+ 2 ( \gamma -1 ) s {\dds} 
+  (  q _1  - 2 q_2 +  \gamma    - 2  )  \varphi  {\ddphi} 
\end{equation}
provided that 
\begin{equation*}   
q_3 = {   q_1 + 2 ( \gamma  -2)  q_2 + \gamma  -  2  \over 2 ( \gamma -1 )} , 
\qquad 
q_4 = {   q_1 + 2 ( \gamma -2)   q_2  +  2 \gamma  - 3  \over 2 ( \gamma -1 )}   .
\end{equation*}

\item 

Case   $ F_0 = 0  $,    $ G_0 \neq 0 $  (any $R_0$)

There exists the symmetry 
\begin{equation}      \label{symmetry_case_3_2} 
X_{2}  
= ( 2 q_1 -   2  \gamma q_3  + \gamma  - 2 )    t  {\ddt} 
+ 2 ( \gamma - 2  ) s {\dds} 
+  (  q_1 -   2 q_3  + \gamma  - 2  )   \varphi  {\ddphi} 
\end{equation}
if 
\begin{equation*}   
q_4 = q_3  +  {  1 \over 2 }  . 
\end{equation*}
The parameter $ q_2 $  is unnecessary in this  case:  
$ F (s) \equiv 0$  because $ F_0 = 0  $.

\end{itemize}

\medskip

\noindent 
{\bf  d)   Exponential functions  case~(\ref{case_exponential}) }

\begin{itemize}

\item

Case 
$ F_0  \neq 0    $  (any $G_0$   and any $R_0$)

There is the symmetry 
\begin{equation}        \label{symmetry_case_4_1} 
X_{2}  
= ( q_1 -  2 \gamma  q_2 )    t  {\ddt} 
+ 2 ( \gamma -1 )  {\dds} 
+  (  q _1-   2 q_2  )  \varphi  {\ddphi} 
\end{equation}
provided that 
\begin{equation*}   
q_3 = q_4 = { q_1 +  2 (  \gamma  - 2  )  q_2   \over 2 ( \gamma -1 )}  . 
\end{equation*}
Requirement~(\ref{condition_q}) 
takes the form $ q_1 ^2 + q_2 ^2   \neq 0  $.

\item 

Case   $ F_0  = 0    $,  $ G_0  \neq 0    $    (any $R_0$)

The PDE admits the additional symmetry  
\begin{equation}      \label{symmetry_case_4_2} 
X_{2}  
= 2 ( q_1 -   \gamma  q_3 )    t  {\ddt} 
+ 2 ( \gamma - 2  )  {\dds} 
+  (  q _1 -   2 q_3  )  \varphi  {\ddphi} 
\end{equation}
with  
\begin{equation*}   
q_4 = q_3   .
\end{equation*}
The parameter $ q_2 $  is unnecessary since  $ F_0 = 0  $.  
Requirement~(\ref{condition_q})  leads to  $ q_1 ^2 + q_3 ^2   \neq 0  $.

\end{itemize}

\subsubsection{Conservation laws in the case $ \gamma \neq 2$ ($\gamma > 1 $)}  

\label{Conservation_laws_not_2}

\medskip

\noindent
{\bf  a)  Arbitrary functions case~(\ref{case_arbitrary})}

The symmetry 
\begin{equation}
Z_1 = {\ddt} 
\end{equation}
is variational. 
It provides the conservation of energy. 
This conservation law also exists in the case of finite conductivity: see~(\ref{CL_A0_energy}). 
Here we  obtain its particular case with $ E^{\theta} = E^{z} \equiv  0$, 
namely 
\begin{multline}   \label{CL_variational_A_energy} 
D_t ^L 
\left\{
{ 1   \over 2 }    (    u  ^2 + v ^2   ) 
 +       
  {  S \over  \gamma  -1  } {   \rho  ^{ \gamma -1}  }
+       
 {   ( H^{\theta}  ) ^2  +  ( H^{z}  ) ^2   \over  2  \rho }
\right\} 
\\
+ 
D_{s}
\left\{
      r  u  \left(  
  p   
  + 
    {  ( H^{\theta}  ) ^2  +  ( H^{z}  ) ^2   \over 2} 
 \right) 
\right\} 
=  0  . 
\end{multline}

The conservation laws for mass~(\ref{Variational_system2_rho}), 
angular momentum (first equation in~(\ref{Variational_system2_v})),  
entropy~(\ref{Variational_system2_S}) 
and magnetic fluxes~(\ref{Variational_system2_H_theta}), (\ref{Variational_system2_H_z})
were  used to obtain the variational formulation of the equations.     
These conservation laws cannot be derived from the Lagrangian. 
In the case of finite conductivity conservation laws of 
mass, angular momentum and magnetic fluxes 
were given by~(\ref{CL_A0_mass}),  
(\ref{CL_A0_rotation}), 
(\ref{CL_A0_flux_theta})  
and~(\ref{CL_A0_flux_z}), respectively. 
The conservation of entropy~(\ref{Modification3_entropy})  
exists only for infinite conductivity.

Thus, for system~(\ref{Variational_system2}) 
we have the same conservation laws   
as for the system~(\ref{Lagrange_system2}) with finite conductivity and 
conservation of entropy. 

\medskip

\noindent
{\bf Special cases of  $ S(s)  $,       $ F(s) $,     $ G(s) $  and    $ R(s)  $  }

Now we examine particular cases of 
\begin{equation}    \label{SFGR} 
 S(s) , 
\qquad 
F(s) = { H^{\theta}  \over  r \rho }  , 
\qquad
G(s)   =  { H^z   \over    \rho  }  , 
\qquad 
R(s)   = r  v    
\end{equation}
with additional symmetries of the Lagrangian. 
We remark that these quantities are particular cases of 
conservation law~(\ref{CL_special_2a}). 
For all cases considered further the additional symmetries turn out to  be variational.

Only additional symmetries and conservation laws,  
i.e. those which exist in addition to  case with arbitrary  
$ S(s)  $,       $ F(s) $,     $ G(s) $  and    $ R(s)  $, 
 are given.

\medskip

\noindent
{\bf b)   Constant functions case~(\ref{case_constant})}


For all constant values  $ S _0  $,       $ F _0  $,     $ G _0  $  and    $ R_0   $ 
there is the symmetry 
\begin{equation}
Z_2 = {\dds} , 
\end{equation}
which is variational. 
It provides the conservation law 
\begin{equation*}      
D_t ^L  \left( 
  \varphi  _{t}     \varphi  _{s}
\right)    
 + 
D_s
\left( 
  -    { 1\over 2 }     \varphi  _{t}   ^2   
+     { 1\over 2 }    { R ^2  \over    \varphi ^2 }
 +     { \gamma  S   \over      \gamma  -1 }     \varphi    ^{1 -  \gamma}       \varphi  _{s}  ^{1 -  \gamma}
 +    {  F   ^2   \varphi    \over     \varphi  _s }  
+    {  G   ^2   \over       \varphi        \varphi  _s  }     
\right) 
= 0 , 
\end{equation*}
which takes the form 
\begin{equation}       \label{conservation_A_s_physical} 
D_t ^L  
\left(  
   {    u   \over r \rho  }     
\right)   
 + 
D_s
\left( 
   {    1  \over 2 }     (  - u  ^2  + v ^2   ) 
 +         {  \gamma  S    \over    \gamma  -1  }  \rho  ^{\gamma -1} 
 +          {  ( H^{\theta}  ) ^2   +   ( H^z ) ^2   \over    \rho }  
\right)  
= 0 
\end{equation}
in the physical variables.

There is one case with specified constants: 

\begin{itemize} 

\item 

{Case}  $\gamma =  { 5 \over 4 } $, $ F_0 \neq 0 $,   $ G_0 = 0 $,  $ R_0  = 0 $

In addition to the symmetries which exist in the base case 
there is the symmetry 
\begin{equation}
Z_{3}  
= 4  t  {\ddt} 
- 2   s {\dds} 
+ 3   \varphi  {\ddphi}   .
\end{equation}

It provides the conservation law
\begin{multline*}
D_t ^L 
\left\{ 
   4  t 
\left(
   {  \varphi_{t}^{2}      \over 2 }
+  {S   \over  \gamma - 1 }     \varphi^{1-\gamma}  \varphi_{s}^{1-\gamma}
+      {  F   ^2   \varphi    \over   2   \varphi  _s }  
 \right)
  - 
2   s  
 \varphi_{s} \varphi_{t} 
 - 
3   \varphi    
 \varphi_{t}   
\right\}
\\
+ 
D_s 
\left\{
 (  4  t   \varphi_{t}   
 -   
 3   \varphi   )
 \left(  S     \varphi^{1-\gamma}  \varphi_{s}^{-\gamma}
+      {  F   ^2   \varphi    \over   2   \varphi  _s ^2 }  
  \right)    
\right. 
\\
\left. 
  -    
2   s  
\left(
 -   { \varphi_{t}^{2}   \over 2 }
+   { \gamma S \over  \gamma - 1 }       \varphi^{1-\gamma}  \varphi_{s}^{1-\gamma}
+      {  F   ^2   \varphi    \over     \varphi  _s }  
 \right) 
\right\} = 0  . 
\end{multline*}
This conservation law gets rewritten in the physical variables as 
\begin{multline}
D_t ^L 
\left\{ 
 4  t 
\left(
   {  u ^{2}  \over 2 }
+  {S  \over  \gamma - 1 }     \rho  ^{\gamma - 1 }
+           {  ( H^{\theta}  ) ^2   \over  2   \rho }
 \right)
 -  
 2   s  
   { u \over r  \rho }   
 -
 3    r   
 u    \right\}
\\
+ 
D_s 
\left\{
   (   4  t  r    u    
- 
 3   r^2 ) 
\left(  
 S   \rho ^{\gamma}    
+     {  ( H^{\theta}  ) ^2     \over  2   }   \right) 
\right. 
\\
\left. 
  -  
2   s  
\left(
 -   { u ^{2}    \over 2 }
+   { \gamma  S \over  \gamma - 1 }     \rho ^{\gamma -1}
+       {  ( H^{\theta}  ) ^2   \over    \rho }  
 \right)  
\right\} = 0  . 
\end{multline}
We remark that this case requires that 
$  H ^z  \equiv 0 $  and $  v \equiv  0$. 
 
\end{itemize}


\medskip

\noindent
{\bf  c)  Power functions case~(\ref{case_power})  }

\begin{itemize}

\item 

{Case}  
$ F_0 \neq 0 $ (any $ G_0 $ and any $ R_0 $)

There is the variational symmetry  
\begin{equation}
Z_{2}  
=  4  (  q_2 +  1  )    t  {\ddt} 
-  2  s {\dds} 
+  ( 2 q _2  + 3  )  \varphi  {\ddphi} 
\end{equation}
provided that 
\begin{equation*}   
q_1 =   - 2  ( \gamma -2 ) q_2 - 4 \gamma + 5   , 
\qquad 
q_3 = - { 3  \over 2  } , 
\qquad 
q_4 = -1    . 
\end{equation*}

Using the symmetry, we obtain  the conservation law
\begin{multline*}
D_t ^L  
\left\{ 
  4  (  q_2 +  1  )     t  
\left(
   {  \varphi_{t}^{2}      \over 2 }
+      { 1\over 2 }    { R ^2  \over    \varphi ^2 }
+   {S   \over  \gamma - 1 }     \varphi^{1-\gamma}  \varphi_{s}^{1-\gamma}
+      {  F   ^2   \varphi    \over   2   \varphi  _s }  
+   {  G   ^2   \over   2      \varphi        \varphi  _s  }     
 \right)
\right.
\\
\left. 
 -  
2   s  
  \varphi_{s} \varphi_{t}  
 -  
  ( 2 q _2  + 3  )  \varphi    
   \varphi_{t}    
\right\}
\\
+ 
D_s 
\left\{
  (  4  (  q_2 +  1  )    t    \varphi_{t}  
-   
  ( 2 q _2  + 3  ) \varphi    ) 
 \left(  S     \varphi^{1-\gamma}  \varphi_{s}^{-\gamma}
+      {  F   ^2   \varphi    \over   2   \varphi  _s ^2 }  
+    {  G   ^2   \over   2      \varphi        \varphi  _s  ^2 }     
  \right)   
\right. 
\\
\left. 
 -  
2  s
\left(
  -  { \varphi_{t}^{2}   \over 2 }
+     { 1\over 2 }    { R ^2  \over    \varphi ^2 }
+   { \gamma S \over  \gamma - 1 }       \varphi^{1-\gamma}  \varphi_{s}^{1-\gamma}
+      {  F   ^2   \varphi    \over     \varphi  _s }  
+   {  G   ^2   \over       \varphi        \varphi  _s  }    
 \right)
\right\} = 0  . 
\end{multline*}

In the physical variables we rewrite it as 
\begin{multline}     \label{CL_power_A_2_1}
D_t ^L 
\left\{ 
  4  (  q_2 +  1  )     t 
\left(
   {  u ^{2}  +  v ^2  \over 2 }
+   {S  \over  \gamma - 1 }     \rho  ^{\gamma - 1 }
+          {  ( H^{\theta}  ) ^2   +   ( H^z ) ^2   \over  2   \rho }
 \right)
\right. 
\\
\left. 
 -  
2  s
   { u \over r  \rho }  
 -  
  ( 2 q _2  + 3  )  r 
   u   
\right\} 
\\
+ 
D_s 
\left\{ 
  ( 4  (  q_2 +  1  )   t    r u   
- 
   ( 2 q _2  + 3  )  r ^2    ) 
\left( 
   S   \rho ^{\gamma}    
+     {  ( H^{\theta}  ) ^2   +   ( H^z ) ^2   \over  2   }   
\right)
\right. 
\\
\left. 
 -  
2  s
\left(
   { - u ^{2}  +  v ^2    \over 2 }
+   { \gamma  S \over  \gamma - 1 }     \rho ^{\gamma -1}
+        {  ( H^{\theta}  ) ^2   +   ( H^z ) ^2   \over    \rho }  
 \right)
\right\}= 0   . 
\end{multline}


\item 

{Case}  $ F_0 = 0  $,    $ G_0 \neq 0 $   (any $ R_0 $) 

There exists the variational symmetry 
\begin{equation}
Z_{2}  
=  2  (   q_1  + 2 \gamma   - 1  )    t  {\ddt} 
+ 2 ( \gamma - 2 ) s {\dds} 
+  (   q_1  +  \gamma  + 1  )  \varphi  {\ddphi} 
\end{equation}
provided that 
\begin{equation*}   
q_3 = - { 3  \over 2  } , 
\qquad 
q_4 = -1  . 
\end{equation*}
Note that there are no restrictions for parameter $q_1$ 
and that the parameter $q_2$ is unnecessary because of   $ F_0 = 0  $.

We obtain the conservation law
\begin{multline*}
D_t ^L  
\left\{ 
   2  (  q_1  + 2 \gamma   - 1  )    t  
\left(
   {  \varphi_{t}^{2}      \over 2 }
+      { 1\over 2 }    { R ^2  \over    \varphi ^2 }
+   {S   \over  \gamma - 1 }     \varphi^{1-\gamma}  \varphi_{s}^{1-\gamma}
+    {  G   ^2   \over   2      \varphi        \varphi  _s  }     
 \right)
\right.
\\
\left. 
  + 
    2 ( \gamma - 2 ) s  
  \varphi_{s} \varphi_{t}  
 -  
  ( q_1 +  \gamma  + 1  )  \varphi    
   \varphi_{t}    
\right\}
\\
+ 
D_s 
\left\{
 (   2  (  q_1 +  2 \gamma - 1  )    t     \varphi_{t}
- 
 ( q_1 + \gamma  + 1  )  \varphi  ) 
  \left(  S     \varphi^{1-\gamma}  \varphi_{s}^{-\gamma}
+    {  G   ^2   \over   2      \varphi        \varphi  _s  ^2 }     
  \right)   
\right. 
\\
\left. 
  +   2 ( \gamma - 2 ) s  
\left(
 -   { \varphi_{t}^{2}   \over 2 }
+       { 1\over 2 }    { R ^2  \over    \varphi ^2 }
+   { \gamma S \over  \gamma - 1 }       \varphi^{1-\gamma}  \varphi_{s}^{1-\gamma}
+    {  G   ^2   \over       \varphi        \varphi  _s  }    
 \right)
\right\} = 0  . 
\end{multline*}

In the physical  variables this conservation law takes the form 
\begin{multline}
D_t ^L 
\left\{ 
  2  (  q_1 + 2 \gamma  - 1  )    t 
\left(
  {  u ^{2}  +  v ^2  \over 2 }
+   {S  \over  \gamma - 1 }     \rho  ^{\gamma - 1 }
+           {   ( H^z ) ^2   \over  2   \rho }
 \right)
\right. 
\\
\left. 
  + 
  2 ( \gamma - 2 ) s   
   { u \over r  \rho }  
- 
 ( q_1 + \gamma   + 1  )   r   
   u    
\right\} 
\\
+ 
D_s 
\left\{ 
  (    2   (  q_1 + 2 \gamma   - 1  )    t   r u    
- 
  (   q_1 +  \gamma + 1  )   r^2  ) 
\left(  
  S   \rho ^{\gamma}    
+    {  ( H^z ) ^2   \over  2   }   \right) 
\right. 
\\
  +  
\left. 
 2 ( \gamma - 2 ) s  
\left(
   { - u ^{2}  +  v ^2    \over 2 }
+  { \gamma  S \over  \gamma - 1 }     \rho ^{\gamma -1}
+       {     ( H^z ) ^2   \over    \rho }  
 \right)
\right\}= 0  . 
\end{multline}
Note that $ H ^{\theta} \equiv 0 $ because of  $ F _ 0 =  0$.  

\end{itemize}


\medskip 

\noindent 
{\bf  d)   Exponential functions  case~(\ref{case_exponential}) }

\begin{itemize}

\item 

{Case} $ F_0 \neq 0 $    (any  $ G_0 $ and  any  $ R_0 $)     

There exists the variational symmetry 
\begin{equation}
Z_{2}  
= 2  q_2     t  {\ddt} 
-  {\dds} 
+  q _2  \varphi  {\ddphi} 
\end{equation}
provided that 
\begin{equation*}   
q_1 = - {   2 ( \gamma - 2  )  q_2  } , 
\qquad 
q_3 = q_4 =  0  . 
\end{equation*}
We requite  $ q_2  \neq 0  $ to ensure~(\ref{condition_q}).

The symmetry provides the  conservation law
\begin{multline*}
D_ t ^L 
\left\{ 
  2  q_2   t   
\left(
   {  \varphi_{t}^{2}      \over 2 }
+      { 1\over 2 }    { R ^2  \over    \varphi ^2 }
+   {S   \over  \gamma - 1 }     \varphi^{1-\gamma}  \varphi_{s}^{1-\gamma}
+      {  F   ^2   \varphi    \over   2   \varphi  _s }  
+    {  G   ^2   \over   2      \varphi        \varphi  _s  }     
 \right)
 - 
   \varphi_{s} \varphi_{t}  
 -   
  q _2  \varphi    
  \varphi_{t}   \right\}
\\
+ 
D_s 
\left\{ 
  q_2  (   2     t    \varphi_{t}
- 
   \varphi    ) 
  \left(  S     \varphi^{1-\gamma}  \varphi_{s}^{-\gamma}
+      {  F   ^2   \varphi    \over   2   \varphi  _s ^2 }  
+    {  G   ^2   \over   2      \varphi        \varphi  _s  ^2 }     
  \right)    
\right. 
\\
\left. 
  - 
\left(
-    { \varphi_{t}^{2}   \over 2 }
+     { 1\over 2 }    { R ^2  \over    \varphi ^2 }
+   { \gamma S \over  \gamma - 1 }       \varphi^{1-\gamma}  \varphi_{s}^{1-\gamma}
+      {  F   ^2   \varphi    \over     \varphi  _s }  
+    {  G   ^2   \over       \varphi        \varphi  _s  }    
 \right)
\right\}  = 0  . 
\end{multline*}

In the physical variables we present it as follows 
\begin{multline}       \label{CL_power_A_3_1}
D_t ^L 
\left\{
    2  q_2     t   
\left(
   {  u ^{2}  +  v ^2  \over 2 }
+   {S  \over  \gamma - 1 }     \rho  ^{\gamma - 1 }
+          {  ( H^{\theta}  ) ^2   +   ( H^z ) ^2   \over  2   \rho }
 \right)
 - 
  { u \over r  \rho }   
-  
  q _2  r 
  u    \right\}
\\
+ 
D_s 
\left\{  
 q_2   (  2    t  r   u  
-   
  r^2 ) 
  \left(  
 S   \rho ^{\gamma}    
+   {  ( H^{\theta}  ) ^2   +   ( H^z ) ^2   \over  2   }   \right) 
\right.
\\ 
\left. 
 - 
\left(
   { - u ^{2}  + v ^2    \over 2 }
+  { \gamma  S \over  \gamma - 1 }     \rho ^{\gamma -1}
+       {  ( H^{\theta}  ) ^2   +   ( H^z ) ^2   \over    \rho }  
 \right) 
\right\}= 0 . 
\end{multline}


\item

{Case} $ F_0 = 0 $,   $ G_0  \neq  0 $   (any  $ R_0 $)  

In  this case $  H ^{\theta}   \equiv 0 $.  
The PDE has the variational symmetry 
\begin{equation}
Z_{2}  
= 2  q_1     t  {\ddt} 
+ 2  (\gamma - 2 )   {\dds} 
+  q _1  \varphi  {\ddphi} 
\end{equation}
provided that 
\begin{equation*}   
q_3 = q_4 =  0  . 
\end{equation*}
We need  $ q_1 \neq 0 $ to keep~(\ref{condition_q}) satisfied.

The symmetry  allows to find the conservation law
\begin{multline*}
D_ t ^L 
\left\{ 
   2  q_1     t 
\left(
   {  \varphi_{t}^{2}      \over 2 }
+      { 1\over 2 }    { R ^2  \over    \varphi ^2 }
+  {S   \over  \gamma - 1 }     \varphi^{1-\gamma}  \varphi_{s}^{1-\gamma}
+    {  G   ^2   \over   2      \varphi        \varphi  _s  }     
 \right) 
  + 
   2  (\gamma - 2 )   
 \varphi_{s} \varphi_{t}  
 -   q _1  \varphi    
     \varphi_{t}  
\right\}
\\
+ 
D_s 
\left\{
 q_1    (     2  t   \varphi_{t} 
- 
  \varphi     ) 
  \left(  S     \varphi^{1-\gamma}  \varphi_{s}^{-\gamma}
+    {  G   ^2   \over   2      \varphi        \varphi  _s  ^2 }     
  \right)    
\right.
\\
\left. 
  + 
  2  (\gamma - 2 )   
\left(
 -   { \varphi_{t}^{2}   \over 2 }
+      { 1\over 2 }    { R ^2  \over    \varphi ^2 }
+   { \gamma S \over  \gamma - 1 }       \varphi^{1-\gamma}  \varphi_{s}^{1-\gamma}
+    {  G   ^2   \over       \varphi        \varphi  _s  }    
 \right)
\right\} = 0  . 
\end{multline*}

This conservation law gets rewritten in the physical  variables  as   
\begin{multline}
D_t  ^L
\left\{  
    2  q_1     t   
\left(
  {  u ^{2}  +  v ^2  \over 2 }
+   {S  \over  \gamma - 1 }     \rho  ^{\gamma - 1 }
+           {     ( H^z ) ^2   \over  2   \rho }
 \right) 
  +    
2  (\gamma - 2 )  
 { u \over r  \rho }  
-  
q _1   r      u   
\right\} 
\\
+ D_s 
\left\{ 
q_1  (     2      t   r  u  
- 
  r   ^2  ) 
  \left(  
  S   \rho ^{\gamma}    
+      {  ( H^z ) ^2   \over  2   }   \right) 
\right.
\\
\left. 
  +  
  2  (\gamma - 2 )  
\left(
   { - u ^{2}  +  v ^2    \over 2 }
+   { \gamma  S \over  \gamma - 1 }     \rho ^{\gamma -1}
+        {     ( H^z ) ^2   \over    \rho }  
 \right)
\right\} = 0 . 
\end{multline}
\end{itemize}


\subsubsection{Symmetries in the case  $\gamma = 2 $}

For $ \gamma =2 $  we examine PDE~(\ref{variational_PDE_gamma}) 
and Lagrangian~(\ref{variational_L_gamma}). 
The analysis is very similar to that of the case $ \gamma  \neq 2 $.  
Symmetries of PDE~(\ref{variational_PDE_gamma}) 
can be given as 
\begin{equation}       \label{generator_form_2_gamma}
X =\sum  _{i=1} ^6  k_i Y_i   , 
\end{equation}
where 
\begin{equation}   
Y_1 =  {\ddt} , 
\quad 
Y_2 =  {\dds} , 
\quad 
Y_3 =   t {\ddt}  , 
\quad 
Y_4 =  s {\dds}  , 
\quad 
Y_5 =   \varphi  {\ddphi}  , 
\quad 
Y_6 =   t ^2 {\ddt}   + t  \varphi  {\ddphi} . 
\end{equation}

Application of operator~(\ref{generator_form_2_gamma}) 
to PDE~(\ref{variational_PDE_gamma})  provides the conditions
\begin{subequations}     \label{conditions_2_gamma}
\begin{gather}
( k_4 s + k_2 )  \tilde{S}_s  (s) 
=  ( -2 k_3  -  k_4 + 4  k_5 )  \tilde{S}  (s)     ,
\label{conditions_2a_gamma}
\\
( k_4 s + k_2 )  F_s (s) 
=  \left( - k_3 -  { 1\over 2 }   k_4 +  k_5 \right)  F (s)  , 
\label{conditions_2b_gamma} 
\\
( k_4 s + k_2 )  R_s (s) 
=  \left( - k_3  +  2 k_5 \right)  R  (s)  , 
\label{conditions_2d_gamma}
\\
k_6 F (s) = 0 
\label{conditions_2f_gamma} 
\end{gather}
\end{subequations}
for coefficients $ k_1, \ldots, k_6$.

Equations~(\ref{conditions_2a_gamma}),  
(\ref{conditions_2b_gamma}) 
and~(\ref{conditions_2d_gamma})   
provide for functions  $ \tilde{S} $, $ F $ and $ R $   
the same cases 
as the   classifying equation~(\ref{classifying_form}). 
Therefore, we get four cases  for each of these functions: 
arbitrary, constant, power and exponential. 
Since not all possible combinations lead to additional symmetries 
we have only four combinations to examine:

\begin{enumerate}

\item

Arbitrary  functions 
\begin{equation}      \label{case_arbitrary_gamma} 
\tilde{S} (s) , 
\quad 
F(s) , 
\quad 
R(s) ;
\end{equation}

\item

Constant functions  
\begin{equation}    \label{case_constant_gamma} 
\tilde{S} (s) = \tilde{S} _{0} , 
\quad 
F(s) =F_{0} , 
\quad 
R(s) =R_{0}  ;
\end{equation}

\item

Power functions  
\begin{equation}      \label{case_power_gamma} 
\tilde{S} (s) = \tilde{S} _{0} s^{q_1}  , 
\quad 
F (s) = F_{0} s^{q_2}  , 
\quad  
R (s) =R_{0} s^{q_3} ; 
\end{equation}

\item

Exponential functions 
\begin{equation}      \label{case_exponential_gamma} 
\tilde{S} (s) = \tilde{S}_{0} e^{q_1 s}  ,  
\quad  
F (s) = F_{0} e^{q_2 s} , 
\quad
R (s) =R_{0} e^{q_3 s}  .
\end{equation}

\end{enumerate}
Here $ \tilde{S}_{0} $, $ F_{0} $ and $ R_{0} $  are constants.  
It should be mentioned that specification of  $ \tilde{S} (s) $ allows 
some freedom  how it can be composed by $ S(s) $ and  $G (s) $ 
(see~(\ref{define_tilde_S})). 
We keep the requirements~(\ref{conditions_S_F_G}). 
Therefore, $  \tilde{S} (s)  \ {\not \equiv } \ 0  $. 
When  $ F (s) \equiv 0$,  we need  $ G(s) \ {\not \equiv } \ 0  $.  
For the last two cases we also need  
\begin{equation}    \label{condition_q_gamma} 
q_1 ^2 + q_2 ^2 +  q_3 ^2   \neq 0 . 
\end{equation}

Now we go through  the  particular cases.


\medskip

\noindent
{\bf  a)  Arbitrary functions case~(\ref{case_arbitrary_gamma})}

For any $ \tilde{S} (s)  $,       $ F(s) $  and    $ R(s)  $  
there is the symmetry 
\begin{equation}
X_1  = {\ddt}   . 
\end{equation}

There is a case with symmetry extension:  

\begin{itemize}

\item 

{Case} $ F(s) \equiv 0 $  ($  G(s) \ {\not \equiv } \ 0  $, any   $ R(s) $) 

There are  two  additional symmetries 
\begin{equation}
X_{*} = 2 t  {\ddt} + \varphi  {\ddphi}, 
\qquad 
X_{**}  = t ^2   {\ddt} + t \varphi  {\ddphi}   . 
\end{equation}

\end{itemize}

\medskip

\noindent
{\bf Special cases of  $ \tilde{S} (s)  $,       $ F(s) $  and    $ R(s)  $ } 

Below  only symmetries which exist in addition to symmetries of the case with 
arbitrary  $ \tilde{S} (s)  $,       $ F(s) $ and    $ R(s)  $ are given.

\medskip

\noindent
{\bf b)   Constant functions case~(\ref{case_constant_gamma})}

For all values $ \tilde{S} _0   $,       $ F _0  $  and    $ R _0  $  
there exists the symmetry 
\begin{equation}
X_2  = {\dds}   .  
\end{equation}

There are more symmetries for particular values 
of the constants:

\begin{itemize} 

\item 

{Case}  $ F_0 \neq 0 $,     $ R_0  = 0 $,  (any $G (s) $) 

There is the additional symmetry 
\begin{equation}
{X} _{3}  
= -  t  {\ddt} 
+ 2 s {\dds}   . 
\end{equation}
This symmetry  is a particular case of~(\ref{symmetry_case_2_2}). 
However, in the case $ \gamma =2 $ there is no requirement  $ G \equiv  0$, 
which we had in the case $\gamma \neq 2 $.

\item 

{Case}  $ F_0 = 0 $,  $ R_0  = 0 $,    ($G (s) \ {\not \equiv } \  0 $)  

There is the additional symmetry 
\begin{equation}
{X} _{3}  
=  t  {\ddt} 
+ 2 s {\dds}  
+  \varphi  {\ddphi}    . 
\end{equation}
It is same symmetry as we had for $\gamma \neq 2 $.   
However, for $ \gamma =2 $ we have less restrictions: 
$G (s) \ {\not \equiv } \  0 $ 
instead of  $ G_0 \neq 0 $. 
\end{itemize}

\medskip

\noindent
{\bf  c)  Power functions case~(\ref{case_power_gamma})  }

\begin{itemize} 

\item 

Case 
$ F_0 \neq 0 $    (any $R_0$)

There is the additional symmetry 
\begin{equation}     
X_{2}  
= ( q_1 -   4 q_2 - 1 )    t  {\ddt} 
+ 2 s {\dds} 
+  (  q _1  - 2 q_2  )  \varphi  {\ddphi} 
\end{equation}
provided that 
\begin{equation*}   
q_3 = {   q_1 +  1  \over 2 }   .
\end{equation*}
It is a particular case of symmetry~(\ref{symmetry_case_3_1}), 
which corresponds to $\gamma =2$.

\item

{Case} $ F_0 = 0  $   ($ G (s)  \ {\not \equiv }  \  0    $,    any $R_0$)

In addition to the symmetries of the arbitrary functions case with $ F(s) \equiv 0 $  
there exists the symmetry 
\begin{equation}
X_{2}  
= 4 s{\dds} 
+   (q _1 +1)  \varphi  {\ddphi}  
\end{equation}
with 
\begin{equation*}   
q_3 = {  q_1  + 1   \over 2 }  . 
\end{equation*}
There are no conditions for  superfluous  parameter $ q_2$.  

\end{itemize}

\medskip

\noindent 
{\bf  d)   Exponential functions  case~(\ref{case_exponential_gamma}) }

\begin{itemize} 

\item

Case 
$ F_0  \neq 0    $  (any $R_0$)

There is the symmetry 
\begin{equation}      
X_{2}  
= ( q_1 -  4 q_2 )    t  {\ddt} 
+ 2 {\dds} 
+  (  q _1-   2 q_2  )  \varphi  {\ddphi} 
\end{equation}
provided that 
\begin{equation*}   
q_3 ={ q_1   \over 2 }  . 
\end{equation*}
Requirement~(\ref{condition_q}) 
takes the form $ q_1  ^2 +   q_2  ^2   \neq 0  $. 
Note that this symmetry is a particular case of~(\ref{symmetry_case_4_1})

\item 

{Case}  $ F_0  = 0    $  ($ G (s)  \   {\not \equiv } \   0    $,    any $R_0$)

In addition to the symmetries of the case of arbitrary functions  with $ F(s) \equiv 0 $  
the PDE admits the symmetry 
\begin{equation} 
X_{2}   
= 4 {\dds} 
+   q _1   \varphi  {\ddphi} 
\end{equation}
for 
\begin{equation*}   
q_3 = { q_1 \over 2} . 
\end{equation*}
Here we need  $ q_1   \neq 0 $.  
The parameter   $ q_2 $  is superfluous.

\end{itemize}


\subsubsection{Conservation laws in the case  $\gamma = 2 $}

The conservation laws which were derived 
for $ \gamma \neq 2 $  in point~\ref{Conservation_laws_not_2} 
also hold for $ \gamma = 2 $. 
Here we refer to point~\ref{Conservation_laws_not_2}  for common conservation laws 
and provide additional    conservation laws, 
which hold only for $ \gamma = 2 $.

\medskip

\noindent
{\bf  a)  Arbitrary functions case~(\ref{case_arbitrary_gamma})}

For all arbitrary functions  $ \tilde{S} (s)  $,       $ F(s) $  and    $ R(s)  $  
we get the same results as for $ \gamma \neq 2 $. 
The variational symmetry 
\begin{equation}
Z_1  = {\ddt}   . 
\end{equation}
provides the conservation of energy~(\ref{CL_variational_A_energy}).   
The conservation laws for mass~(\ref{Variational_system2_rho}), 
angular momentum (first equation in~(\ref{Variational_system2_v})),  
entropy~(\ref{Variational_system2_S}) 
and magnetic fluxes~(\ref{Variational_system2_H_theta}),  (\ref{Variational_system2_H_z})
cannot be derived from the Lagrangian 
because they were used to bring the equation into the  variational form.

There is a case with additional conservation laws:

\begin{itemize} 

\item 

{Case}   $ F(s) \equiv 0 $ ($  G(s) \ {\not \equiv } \ 0  $,  any   $ R(s) $)

There is  an additional  variational symmetry
\begin{equation}     \label{symmetry_gamma_1}
Z_{*}  
= 2 t  {\ddt} + \varphi  {\ddphi}, 
\end{equation}
and  an  additional  divergence  symmetry 
\begin{equation}     \label{symmetry_gamma_2}
Z_{**}  
 = t ^2   {\ddt} + t \varphi  {\ddphi}   
\qquad 
\mbox{with} 
\qquad 
 ( B_1 , B_2 ) = \left( { \varphi ^2 \over 2 } , 0 \right)  . 
\end{equation}

The first additional symmetry   $ Z_{*}  $ 
provides the conservation law
\begin{multline*}
D_t ^L  
\left\{
    2 t
\left(
  {  \varphi_{t}^{2}      \over 2 }
+      { 1\over 2 }    { R ^2  \over    \varphi ^2 }
+ {   \tilde{S}     \over  \gamma - 1 }     \varphi^{1-\gamma}  \varphi_{s}^{1-\gamma}
 \right)
  -   {\varphi}    
 \varphi_{t}  
\right\} 
\\
+ 
D_s
\left\{
   (  2 t    \varphi_{t}  
-  
 {\varphi}     )  
\tilde{S}     \varphi^{1-\gamma}  \varphi_{s}^{-\gamma}
\right\} 
= 0  . 
\end{multline*}

This conservation law can be rewritten in  physical variables as 
\begin{multline}     \label{CL_gamma_1} 
D_t ^L  
\left\{
   2 t 
\left(
  {  u ^{2}  +  v ^2  \over 2 }
+   {S  \over  \gamma - 1 }     \rho  ^{\gamma - 1 }
+          {    ( H^z ) ^2   \over  2   \rho }
 \right)
  -   r 
    u   
\right\} 
\\
+ 
D_s
\left\{
  (  2 t  r    u 
- 
 r ^2 ) 
  \left(  
  S   \rho ^{\gamma}    
+     {    ( H^z ) ^2   \over  2   }   \right) 
\right\} 
= 0   . 
\end{multline}


The second additional symmetry  $ Z_{**}   $ 
leads to conservation law
\begin{multline*}
D_t ^L  
\left\{ 
     t   ^2 
\left( 
 {  \varphi_{t}^{2}      \over 2 }
+      { 1\over 2 }    { R ^2  \over    \varphi ^2 }
+   {  \tilde{S}    \over  \gamma - 1 }     \varphi^{1-\gamma}  \varphi_{s}^{1-\gamma}
 \right) 
  - 
 t  {\varphi}    \varphi_{t}   
+ 
 { \varphi ^2 \over 2 }
\right\} 
\\
+ 
D_s
\left\{
 (  t^2   \varphi_{t} 
- 
 t  {\varphi}   ) 
\tilde{S}     \varphi^{1-\gamma}  \varphi_{s}^{-\gamma}
\right\} 
= 0 , 
\end{multline*}
which can be rewritten in the physical variables as follows 
\begin{multline}      \label{CL_gamma_2} 
D_t ^L  
\left\{
   t ^2 
\left(
   {  u ^{2}  +  v ^2  \over 2 }
+  {S  \over  \gamma - 1 }     \rho  ^{\gamma - 1 }
+           {     ( H^z ) ^2   \over  2   \rho }
 \right)
- 
 t  r    u    
+ 
 { r  ^2 \over 2 } 
\right\} 
\\
+ 
D_s
\left\{
 (   t ^2    r      u 
  -  
 t    r^2   ) 
   \left(  
 S   \rho ^{\gamma}    
+     {     ( H^z ) ^2   \over  2   }   \right) 
\right\} 
= 0  . 
\end{multline}
We note that for conservation laws~(\ref{CL_gamma_1})  and~(\ref{CL_gamma_2}) 
$ H^{\theta} \equiv 0 $ because of $ F_0 = 0 $.   
\end{itemize}


\medskip

\noindent 
{\bf Special cases of  $ \tilde{S} (s)  $,       $ F(s) $   and    $ R(s) $ } 

Now we examine particular cases of 
\begin{equation}    \label{tildeSFR} 
 \tilde{S} (s) = S(s) + { 1 \over 2 }  \left(   { H^z   \over    \rho  }  \right) ^2  , 
\qquad 
F(s) = { H^{\theta}  \over  r \rho }  , 
\qquad 
R(s)   = r  v    
\end{equation}
with additional symmetries of the Lagrangian. 
We remark that these quantities are particular cases of 
conservation law~(\ref{CL_special_2a}). 
For all cases considered further the additional symmetries are variational.


\medskip

\noindent
{\bf b)   Constant functions case~(\ref{case_constant_gamma})}

For all constant values  $ \tilde{S} _0   $,       $ F _0 $  and    $ R _0  $   
there is variational symmetry 
\begin{equation}   \label{s_translation_gamma} 
Z_2  = {\dds}   . 
\end{equation}
It provides the particular case  of conservation  law~(\ref{conservation_A_s_physical}), 
corresponding to $ \gamma =2 $.

For particular values of the constants we obtain one case. 

\begin{itemize}

\item 

Case $F_0 = 0$    ($  G(s) \ {\not \equiv } \ 0  $,  any   $ R _0  $)

We get the same conservation laws as for arbitrary functions with $ F(s) \equiv 0 $  
and the conservation law  given by symmetry~(\ref{s_translation_gamma}).

\end{itemize}

\medskip

\noindent
{\bf  c)  Power functions case~(\ref{case_power_gamma})  }

\begin{itemize}

\item 

{Case}  $ F_0  \neq  0  $ (any $ R_0$)

There is a  variational symmetry  
\begin{equation}
Z_{2}  
=  4  (  q_2 +  1  )    t  {\ddt} 
-  2  s {\dds} 
+  ( 2 q _2  + 3  )  \varphi  {\ddphi} 
\end{equation}
if 
\begin{equation*}   
q_1  = -3 , 
\qquad 
q_3 = -1    .
\end{equation*}
There are no restrictions for $q_2$.  
We obtain conservation law~(\ref{CL_power_A_2_1}) 
with specified parameter $ \gamma = 2 $.

\item 

{Case}  $ F_0 = 0  $  ($ G (s) \ {\not \equiv } \   0 $, any $ R_0$)    

In addition to the symmetries of the arbitrary function case with $ F(s)  \equiv 0$  
there is the variational symmetry    
\begin{equation}
Z_{2}   
=   t  {\ddt} 
+   s {\dds} 
\end{equation}
for  
\begin{equation*}   
q_1  = -3 , 
\qquad 
q_3 = -1    .
\end{equation*}

Thus,  in addition to the conservation laws  of the arbitrary function case with $ F(s)  \equiv 0$  
we obtain   the conservation law
\begin{multline*}
D _t ^L 
\left\{ 
   t
\left(
  {  \varphi_{t}^{2}      \over 2 }
+       { 1\over 2 }    { R ^2  \over    \varphi ^2 }
+   {  \tilde{S}    }     \varphi^{1-\gamma}  \varphi_{s}^{1-\gamma}
 \right)
  + 
 s
  \varphi_{s} \varphi_{t}  
\right\}
\\
+ 
D_s 
\left\{  
   t
  \varphi_{t}   
    {   \tilde{S}    \over        \varphi        \varphi  _s  ^2 }     
  + 
 s
\left(
  -  { \varphi_{t}^{2}   \over 2 }
+       { 1\over 2 }    { R ^2  \over    \varphi ^2 }
+    {   \tilde{S}    \over       \varphi        \varphi  _s  }    
 \right)
\right\}   =  0  . 
\end{multline*}

In the physical variables it takes the form 
\begin{multline}
D_t  ^L
\left\{ 
   t
\left(
   {  u ^{2}  +  v ^2  \over 2 }
+   {S  \over  \gamma - 1 }     \rho  ^{\gamma - 1 }
+           {     ( H^z ) ^2   \over  2   \rho }
 \right)
  +  
  s
   { u \over r  \rho }   
\right\} 
\\
+ 
 D_s 
\left\{ 
    t
  r   u    \left(  
    S   \rho ^{\gamma}    
+       {     ( H^z ) ^2   \over  2   }   \right) 
  +  s
\left(
   {  - u ^{2}  +  v ^2    \over 2 }
+  { \gamma  S \over  \gamma - 1 }     \rho ^{\gamma -1}
+        {     ( H^z ) ^2   \over    \rho }  
 \right)
\right\} = 0   . 
\end{multline}
In this case $ H ^{ \theta } \equiv 0 $  that follows from $ F_0 = 0 $.

\end{itemize}


\medskip

\noindent 
{\bf  d)   Exponential functions  case~(\ref{case_exponential_gamma}) }

\begin{itemize}

\item 

{Case}  $ F_0  \neq  0  $ (any  $ R_0   $)   

There is the variational symmetry 
\begin{equation}
Z
=  2 q_2  t  {\ddt} 
 -  {\dds} 
+ q_2 {\varphi}  {\ddphi}
\end{equation}
provided that 
\begin{equation*}
q_1 = 0  , 
\qquad 
q_3 = 0 . 
\end{equation*}
We need $ q_2 \neq 0$.  
The symmetry provides the conservation laws~(\ref{CL_power_A_3_1}) 
with specified value $ \gamma = 2 $.

\item 

{Case}  $ F_0 = 0  $ ($ G (s) \   {\not \equiv } \  0 $,  any $ R_0$)    

There are no additional  variational symmetries. 
We get the same conservation laws as in  the case 
of arbitrary functions with  $ F (s)  \equiv 0 $.

\end{itemize}



\section{Concluding remarks}

\label{Concluding}

In the present paper we considered 
Lie point symmetries and  conservation laws  
of the one-dimensional MGD flows 
with cylindrical symmetry. 
Such flows in the mass Lagrangian coordinates are described by equations~(\ref{Lagrange_system}). 
Up to our knowledge these equations were not 
examined for  symmetries and  conservation laws  before. 
The analysis of the equations specifies four different cases 
for the electric conductivity $ \sigma ( \rho , p) $  and constant $ A $, 
which provides the radial  component of the magnetic fields~(\ref{A_value}). 
These cases are

\begin{itemize} 

\item 

finite  electric conductivity  $ \sigma ( \rho , p) $ and  $ H^r = A / r $ with $ A \neq 0$;

\item 

finite  electric conductivity $ \sigma ( \rho , p) $ and $ H^r \equiv 0  $ ($ A=  0$);

\item 

infinite  electric conductivity and  $ H^r = A / r $ with $ A \neq 0$;
 
\item 

infinite  electric conductivity and $ H^r \equiv 0  $ ($ A=  0$). 
     
\end{itemize} 
The last case splits into two {sub}cases for the values  
of the polytropic constant: 
generic   {sub}case $ \gamma \neq 2 $  
and special   {sub}case $ \gamma = 2 $. 
Lie point symmetries were found for all cases.  
In the cases with finite conductivity 
Lie group classifications for $ \sigma ( \rho , p) $  were provided.

The conservation laws were found by different methods. 
Direct computation was used for the cases 
with finite  electric conductivity $ \sigma ( \rho , p) $. 
Variational formulation of the differential equations 
was involved for the cases with 
infinite   electric conductivity. 
It allows to use the Noether theorem for computation of conservation laws. 
For $ A \neq 0 $ we obtained Lie group classification considering 
the entropy function $ S(s)$  as an arbitrary element. 
For   $ A  =  0 $ 
there were four arbitrary elements  
$ S(s)$, $ F(s)$, $ G(s)$ and $ R(s)$,  
given in~(\ref{SFGR}),  
for $ \gamma  \neq 2 $ 
and three arbitrary elements 
$ \tilde{S} (s)$, $ F(s)$ and $ R(s)$,  
provided in~(\ref{tildeSFR}),  
for $ \gamma  = 2 $.  
For both {sub}cases 
$ \gamma  \neq 2 $ and $ \gamma  = 2 $ 
there were provided Lie group classifications. 
The Noether theorem was employed 
for computation of conservation laws.     
The conservation laws, 
obtained by variational approach,   
were also given in the original (physical) variables.

\section*{Acknowledgements}

The research was supported by Russian Science Foundation Grant no. 18-11-00238
"Hydrodynamics-type equations: symmetries, conservation laws, invariant difference
schemes". 
E.I.K. sincerely appreciates the hospitality of the Suranaree University of Technology.


\bigskip

\noindent{\bf \Large Appendices}

\appendix

\section{Lagrangian variables}

\label{Appendix_Lagrangian_variables}

We denote the Eulerian spacial Cartesian coordinates and the velocity components 
as $(x,y,z)$ and $(u^x,u^y,u^z)$, 
the corresponding Lagrangian spacial coordinates 
as $(X,Y,Z)$,
the Eulerian cylindrical coordinates and the velocity components  
as  $(r,\theta,z)$  and  $(u,v,w)$, 
and 
the corresponding Lagrangian cylindrical coordinates 
 as $(\xi,\eta,\zeta)$. 
Relations between Eulerian and Lagrangian coordinates
are as follows
\begin{equation*} 
x=\varphi^{x}(t,X,Y,Z), 
\qquad 
y=\varphi^{y}(t,X,Y,Z),
\qquad 
z=\varphi^{z}(t,X,Y,Z);
\end{equation*}
\begin{equation}
r=\varphi^{r}(t,\xi,\eta,\zeta),
\qquad
\theta=\varphi^{\theta}(t,\xi,\eta,\zeta),
\qquad 
z=\tilde{\varphi}^{z}(t,\xi,\eta,\zeta),
\label{eq:june4.1}
\end{equation}
and
\begin{equation*} 
\varphi^{x}(t_{0},X,Y,Z)=X,
\qquad 
\varphi^{y}(t_{0},X,Y,Z)=Y, 
\qquad 
\varphi^{z}(t_{0},X,Y,Z)=Z;
\end{equation*}
\begin{equation*} 
\varphi^{r}(t_{0},\xi,\eta,\zeta)=\xi,
\qquad 
\varphi^{\theta}(t_{0},\xi,\eta,\zeta)=\eta, 
\qquad 
\tilde{\varphi}^{z}(t_{0},\xi,\eta,\zeta)=\zeta,
\end{equation*}
where the functions $\varphi^{x}(t,X,Y,Z)$, $\varphi^{y}(t,X,Y,Z)$
and $\varphi^{z}(t,X,Y,Z)$ satisfy the Cauchy problem
\begin{equation*} 
\begin{array}{lll}
\varphi_{t}^{x}=u^x(t,\varphi^{x},\varphi^{y},\varphi^{z}), &  &\varphi^{x}(t_{0},X,Y,Z)=X,
\\
\varphi_{t}^{y}=u^y(t,\varphi^{x},\varphi^{y},\varphi^{z}), &  & \varphi^{y}(t_{0},X,Y,Z)=Y,
\\
\varphi_{t}^{z}=u^z(t,\varphi^{x},\varphi^{y},\varphi^{z}), &  & \varphi^{z}(t_{0},X,Y,Z)=Z.
\end{array}
\end{equation*}

 Relations between Cartesian and cylindrical Eulerian coordinates
are
\begin{equation*} 
x=r\cos\theta,
\qquad 
y=r\sin\theta,
\end{equation*}
\begin{equation*} 
u^x=u\cos\theta-v\sin\theta,
\qquad
u^y=u\sin\theta+v\cos\theta  , 
\qquad 
u^z = w . 
\end{equation*}
Because of the relations at $t=t_{0}$, one has that
\begin{equation}
X=\xi\cos\eta, 
\qquad  
Y=\xi\sin\eta, 
\qquad 
Z=\zeta.
\label{eq:june4.2}
\end{equation}
One finds the Jacobians
\begin{equation*} 
\frac{\partial(x,y,z)}{\partial(r,\theta,z)}=\det\left(\begin{array}{ccc}
\cos\theta & -r\sin\theta & 0\\
\sin\theta & r\cos\theta & 0\\
0 & 0 & 1
\end{array}\right)=r 
\end{equation*}
and 
\begin{equation*} 
\frac{\partial(X,Y,Z)}{\partial(\xi,\eta,\zeta)}=\det\left(\begin{array}{ccc}
\cos\eta & -\xi\sin\eta & 0\\
\sin\eta & \xi\cos\eta & 0\\
0 & 0 & 1
\end{array}\right)=\xi.
\end{equation*}

Noticing that
\begin{equation*} 
\dot{x}=u^x, 
\qquad 
\dot{y}=u^y, 
\qquad 
\dot{z}=u^z,
\end{equation*}
where the dot `$\ \dot{}\ $' means the derivative with respect to
$t$, one derives
\begin{equation*} 
\begin{array}{c}
(\dot{r}-u)\cos\theta-(r\,\dot{\theta}-v)\sin\theta=0,\\
(\dot{r}-u)\sin\theta+(r\,\dot{\theta}-v)\cos\theta=0.
\end{array}
\end{equation*}
The latter determines the relations
\begin{equation*} 
\dot{r}=\varphi_{t}^{r}=u,
\qquad 
r\dot{\theta}=\varphi^{r}\varphi_{t}^{\theta}=v,
\end{equation*}
which means that the functions~(\ref{eq:june4.1}) satisfy the Cauchy
problem
\begin{equation}
\begin{array}{lll}
\varphi_{t}^{r}=u(t,\varphi^{r},\varphi^{\theta},\varphi^{\zeta}), 
& & \varphi^{r}(t_{0},\xi,\eta,\zeta)=\xi,
\\
\varphi_{t}^{\theta} 
=  {  \displaystyle v(t,\varphi^{r},\varphi^{\theta},\varphi^{\zeta}) \over  \displaystyle \varphi^{r}  }, 
& & \varphi^{\theta}(t_{0},\xi,\eta,\zeta)=\eta,
\\
\tilde{\varphi}_{t}^{z}=w(t,\varphi^{r},\varphi^{\theta},\varphi^{\zeta}), 
& & \tilde{\varphi}^{z}(t_{0},\xi,\eta,\zeta)=\zeta.
\end{array}\label{eq:june4.3}
\end{equation}

The general solution of the conservation law of mass equation is
\begin{equation*} 
\rho(t,X,Y,Z)=\frac{\rho_{0}(X,Y,Z)}{J(t,X,Y,Z)},
\end{equation*}
where $\rho_{0}(X,Y,Z)$ is some function, and $J$ is the Jacobian
\begin{equation*} 
J=\frac{\partial(\varphi^{x},\varphi^{y},\varphi^{z})}{\partial(X,Y,Z)}.
\end{equation*}
Because of the property 
\begin{equation*} 
J 
=\frac{\partial(\varphi^{x},\varphi^{y},\varphi^{z})}{\partial(X,Y,Z)}
=\frac{\partial(x,y,z)}{\partial(r,\theta,z)} 
\ 
\frac{\partial(r,\theta,z)}{\partial(\xi,\eta,\zeta)}
\ 
\left(\frac{\partial(X,Y,Z)}{\partial(\xi,\eta,\zeta)}\right)^{-1}
=\frac{r}{\xi}J_{0}
\end{equation*}
we obtain\footnote{Here $X$, $Y$ and $Z$ are substituted from~(\ref{eq:june4.2}).}
\begin{equation*} 
\rho(t,\xi,\eta,\zeta)
=\frac{\xi\rho_{0}(\xi,\eta,\zeta)}{\varphi^{r}(t,\xi,\eta,\zeta)J_{0}(t,\xi,\eta,\zeta)},
\end{equation*}
where
\begin{equation*} 
J_{0}=\frac{\partial(r,\theta,z)}{\partial(\xi,\eta,\zeta)}.
\end{equation*}

In the present paper we  assume that all dependent
variables in cylindrical coordinates  depend only on $t$ and $r$. 
In particular,
\begin{equation*} 
u=u(t,r), 
\qquad 
v=v(t,r), 
\qquad 
w=w(t,r),
\qquad 
\rho=\rho(t,r).
\end{equation*}
The latter gives that a solution of the Cauchy problem~(\ref{eq:june4.3})
has the form
\begin{equation*} 
\varphi^{r}=\tilde{\varphi}(t,\xi),
\qquad 
\varphi^{\theta}=\tilde{\psi}(t,\xi)+\eta,
\qquad 
\varphi^{\zeta}=\tilde{\chi}(t,\xi)+\zeta.
\end{equation*}
Hence, 
$J_{0}=\tilde{\varphi}_{\xi}$ and $\rho_{0}=\rho_{0}(\xi).$
Introducing the variable $s=\alpha(\xi)$ such that 
$\alpha^{\prime}(\xi)=\xi\rho_{0}(\xi)$  
and the function $\varphi(t,s)$ such that 
$\tilde{\varphi}(t,\xi)=\varphi(t,\alpha(\xi))$,
one obtains
\begin{equation*} 
\rho(t,s)=\frac{1}{\varphi(t,s)\varphi_{s}(t,s)}.
\end{equation*}
The coordinate $s$ is called the mass Lagrangian coordinate.

Summarizing, one obtains that in the mass Lagrangian coordinates
\begin{equation*} 
\varphi_{t}=u(t,\varphi), 
\qquad 
\varphi_{s}  =\frac{1}{\varphi  \rho }; 
\end{equation*}
\begin{equation*} 
\psi_{t}=   { v(t,\varphi)  \over  \varphi }  ; 
\end{equation*}
\begin{equation*} 
\chi_{t}=w(t,\varphi).
\end{equation*}

\section{Equivalence transformations} 

\label{Equivalence}

\subsection{The case of finite electric conductivity $\sigma (\rho , p) $ and  $ A  \neq 0 $}



Equivalence transformations of the system~(\ref{Lagrange_system}) 
with $  A \neq  0$ have the form 
\begin{multline}    \label{finite_equivalence_A}
X ^e  = \zeta^t \frac{\partial}{\partial{t}}
    + \zeta^s \frac{\partial}{\partial{s}}
    + \zeta^r \frac{\partial}{\partial{r}}
    + \zeta^{\theta} \frac{\partial}{\partial{\theta}}
    + \zeta^z \frac{\partial}{\partial{z}}
    + \zeta^u \frac{\partial}{\partial{u}}
    + \zeta^v \frac{\partial}{\partial{v}}
    + \zeta^w \frac{\partial}{\partial{w}}
    + \zeta^\rho \frac{\partial}{\partial{\rho}}
    + \zeta^p \frac{\partial}{\partial{p}}
\\
    + \zeta^{E^{\theta}} \frac{\partial}{\partial{E^{\theta}}}
    + \zeta^{E^z} \frac{\partial}{\partial{E^z}}
    + \zeta^{H^{\theta}} \frac{\partial}{\partial{H^{\theta}}}
    + \zeta^{H^z} \frac{\partial}{\partial{H^z}}
    + \zeta^{\sigma} \frac{\partial}{\partial{\sigma}}
    + \zeta^{A} \frac{\partial}{\partial{A}},
\end{multline}
where $\zeta^t$, $\zeta^s$,  ... , $\zeta^{\sigma}$,   $\zeta^{A}$    are functions
of $t$, $s$, ${r}$,  $ {\theta} $, $z$,  ${u}$, $v$, $w$, $\rho$, $p$,  
$E^{\theta}$, $E^z$,  $H^{\theta}$, $H^z$, $\sigma$ and $A$.  
Computation leads to the generators
\begin{multline}     \label{equivalence_2}
X ^e _1 =  {\ddt},
\qquad
X ^e _2 =  {\dds},
\\
X ^e _{3} =
    t {\ddt}
    + 2 s {\dds}
    - u  {\ddu }  - v {\ddu} - w {\ddw}
    + 2 \rho {\ddrho}
    - E^{\theta} {\ddEtheta} 
    - E^z {\ddEz}
    + \sigma { \partial \over \partial \sigma } ,
\\  
X ^e _{4} =
    -  2 s {\dds}
    + r  {\ddr}   + z {\ddz}
    + u {\ddu}  + v {\ddv} +  w {\ddw}
    -  4 \rho {\ddrho}      -  2 p {\ddp}  
    -  H^{\theta}  {\ddHtheta}
    -  H^z {\ddHz}
    -  2 \sigma { \partial \over \partial \sigma }  ,
\\  
X ^e _{5} =   2s {\dds}
    + 2\rho {\ddrho}
    + 2p {\ddp}
    + E^{\theta}{\ddEtheta}
    + E^z {\ddEz}
    + H^{\theta} {\ddHtheta}
    + H^z {\ddHz}
    + A { \partial \over \partial {A}  } , 
\\
 X ^e _{6} =   t {\ddz} + {\ddw} ,
\qquad 
X ^e  _7 =   {\phi} _{1}  (s)     {\partial  \over  \partial \theta }   ,   
 \qquad 
 X ^e  _{8} = {\phi} _{2} (s) {\ddz} , 
\end{multline}
where ${\phi} _1 (s) $ and ${\phi}_{2} (s) $    are arbitrary functions.

\subsection{The case of finite electric  conductivity $\sigma (\rho , p) $ and $  A  =  0 $}


The reduced  system~(\ref{Lagrange_system2})
has {equivalence transformations} 
provided by the generators 
\begin{multline}     \label{finite_equivalence_no_A}
X ^e  = \zeta^t \frac{\partial}{\partial{t}}
    + \zeta^s \frac{\partial}{\partial{s}}
    + \zeta^r \frac{\partial}{\partial{r}}
  + \zeta^{\theta} \frac{\partial}{\partial{\theta}}
    + \zeta^u \frac{\partial}{\partial{u}}
    + \zeta^v \frac{\partial}{\partial{v}}
    + \zeta^\rho \frac{\partial}{\partial{\rho}}
    + \zeta^p \frac{\partial}{\partial{p}}
\\
    + \zeta^{E^{\theta}} \frac{\partial}{\partial{E^{\theta}}}
    + \zeta^{E^z} \frac{\partial}{\partial{E^z}}
    + \zeta^{H^{\theta}} \frac{\partial}{\partial{H^{\theta}}}
    + \zeta^{H^z} \frac{\partial}{\partial{H^z}}
    + \zeta^{\sigma} \frac{\partial}{\partial{\sigma}}   , 
\end{multline}
where $\zeta^t$, $\zeta^s$,  ... , $\zeta^{\sigma}$     are functions
of $t$, $s$, ${r}$,  $ {\theta} $,   ${u}$, $v$, $\rho$, $p$,  
$E^{\theta}$, $E^z$,  $H^{\theta}$, $H^z$ and $\sigma$. 
Computations provide   
\begin{multline}     \label{equivalence_4}
X ^e _1 =  {\ddt},
\qquad
X ^e _2 =  {\dds},
\\
X ^e _{3} =
    t {\ddt}
    + 2 s {\dds}
    - u  {\ddu }  - v {\ddu} 
    + 2 \rho {\ddrho}
    - E^{\theta} {\ddEtheta} 
    - E^z {\ddEz}
    + \sigma { \partial \over \partial \sigma } ,
\\  
X ^e _{4} =
    -  2 s {\dds}
    + r  {\ddr}   
    + u {\ddu}  + v {\ddv} 
    -  4 \rho {\ddrho}      -  2 p {\ddp}  
    -  H^{\theta}  {\ddHtheta}
    -  H^z {\ddHz}
    -  2 \sigma { \partial \over \partial \sigma }  ,
\\  
X ^e _{5} =   2s {\dds}
    + 2\rho {\ddrho}
    + 2p {\ddp}
    + E^{\theta}{\ddEtheta}
    + E^z {\ddEz}
    + H^{\theta} {\ddHtheta}
    + H^z {\ddHz}  , 
 \\
X ^e  _6 =   {\phi}   (s , r v )     {\partial  \over  \partial \theta }   ,  
\end{multline}
where ${\phi}  (s , r v ) $ is an arbitrary function.

\subsection{The case of infinite electric conductivity and   $ A \neq  0 $}


The system~(\ref{Lagrange_system3}) 
possesses equivalence transformations given by the generators 
\begin{multline} \label{infinite_equivalence_A}
X ^e  = \zeta^t \frac{\partial}{\partial{t}}
    + \zeta^s \frac{\partial}{\partial{s}}
    + \zeta^r \frac{\partial}{\partial{r}}
    + \zeta^{\theta} \frac{\partial}{\partial{\theta}}
    + \zeta^z \frac{\partial}{\partial{z}}
    + \zeta^u \frac{\partial}{\partial{u}}
    + \zeta^v \frac{\partial}{\partial{v}}
    + \zeta^w \frac{\partial}{\partial{w}}
      \\
    + \zeta^\rho \frac{\partial}{\partial{\rho}}
    + \zeta^p \frac{\partial}{\partial{p}}
    + \zeta^{E^{\theta}} \frac{\partial}{\partial{E^{\theta}}}
    + \zeta^{E^z} \frac{\partial}{\partial{E^z}}
    + \zeta^{H^{\theta}} \frac{\partial}{\partial{H^{\theta}}}
    + \zeta^{H^z} \frac{\partial}{\partial{H^z}}
    + \zeta^{A} \frac{\partial}{\partial{A}},
 \end{multline}
where the coefficients  $\zeta^t$, $\zeta^s$,  ... ,   $\zeta^{A}$    are functions
of $t$, $s$, ${r}$,  $ {\theta} $, $z$,  ${u}$, $v$, $w$, $\rho$, $p$,  
$E^{\theta}$, $E^z$,  $H^{\theta}$, $H^z$ and $A$. 
Computations lead to  the  generators
\begin{multline}   \label{infinite_equivalence_C_generators}
X ^e _1 =  {\ddt},
\qquad
X ^e _2 =  {\dds},
\qquad 
X ^e _{3} =
    t {\ddt}
    + 2 s {\dds}
    - u  {\ddu }  - v {\ddu} - w {\ddw}
    + 2 \rho {\ddrho}  ,
\\
X ^e _{4} =
    -  2 s {\dds}
    + r  {\ddr}   + z {\ddz}
    + u {\ddu}  + v {\ddv} +  w {\ddw}
    -  4 \rho {\ddrho}      -  2 p {\ddp}  
    -  H^{\theta}  {\ddHtheta}
    -  H^z {\ddHz}  ,
\\  
X ^e _{5} =   2s {\dds}
    + 2\rho {\ddrho}
    + 2p {\ddp}
    + H^{\theta} {\ddHtheta}
    + H^z {\ddHz}
    + A { \partial \over \partial {A}  } , 
\\
 X ^e _{6} =   t {\ddz} + {\ddw} ,
\qquad 
X ^e  _7 =   {\phi} _{1}    \left( s,  { p \over \rho^{\gamma}  } \right)   {\partial  \over  \partial \theta }   ,   
 \qquad 
 X ^e  _{8} = {\phi} _{2}  \left( s,  { p \over \rho^{\gamma}  } \right)  {\ddz},
\end{multline}
where $   {\phi} _{1}  $ and $   {\phi} _{2}  $ are arbitrary functions of their arguments.

\subsection{Variational approach
for infinite electric conductivity  and  $ A  \neq  0 $}


The   system  of equations~(\ref{variational_three_PDEs})   
has equivalence transformation of the form 
\begin{equation}
X ^e    = \zeta^t \frac{\partial}{\partial{t}}
    + \zeta^s \frac{\partial}{\partial{s}}
    + \zeta^{\varphi} { \partial  \over \partial \varphi}
    + \zeta^{\psi} { \partial  \over \partial \psi}
    + \zeta^{\chi } { \partial  \over \partial \chi }
    + \zeta^{A} \frac{\partial}{\partial{A}}
    + \zeta^{S} \frac{\partial}{\partial{S}},
\end{equation}
with the coefficients which depend 
on $  t $, $s$,  $\varphi $,   $ \psi$,  $ \chi$,  $A$  and $S$. 
We obtain  the following generators  
\begin{multline}     \label{equivalence_S}
X ^e  _1
=
{\ddt} ,
\qquad
X ^e  _2
=
{\dds} ,
\qquad
X ^e  _3
=
 { \partial  \over \partial \psi} ,
\qquad
X ^e  _4
=
{ \partial  \over \partial \chi  } ,
\qquad
X ^e  _5
=
t { \partial  \over \partial \chi  } ,
\\
X ^e  _{6}
=
  ( 2 \gamma -1)  t {\ddt}
+ 2 ( \gamma -1 ) s  {\dds}
+ \gamma \left(  \varphi   { \partial  \over \partial \varphi}
+  \chi  { \partial  \over \partial \chi  }  \right)   ,
\\
X ^e  _{7}
=   t {\ddt}
+ 2 s  {\dds}
 - 2  \gamma    S  {\partial \over  \partial  S  }    ,
\qquad
X ^e  _{8}
=
( 1 - \gamma)     t {\ddt}
+ 2 s  {\dds}
+  \gamma  A   {\partial \over  \partial  A   } .
\end{multline}


\subsection{Variational approach
for infinite electric conductivity and     $ A  =  0 $}


PDEs~(\ref{variational_PDE}) and~(\ref{variational_PDE_gamma}) 
have equivalence transformations 
with generators of the forms
\begin{equation}    \label{generators_generic}
X^{e}
= \eta^{t} \frac{\partial}{\partial t}
+\eta^{s} \frac{\partial}{\partial s}
+\eta^{\varphi} \frac{\partial}{\partial\varphi}
+\eta^{S} \frac{\partial}{\partial S}
+\eta^{F} \frac{\partial}{\partial F}
+\eta^{G} \frac{\partial}{\partial G}
+\eta^{R} \frac{\partial}{\partial R}
\end{equation}
and
\begin{equation}    \label{generators_gamma}
X^{e}
= \eta^{t} \frac{\partial}{\partial t}
+\eta^{s} \frac{\partial}{\partial s}
+\eta^{\varphi} \frac{\partial}{\partial\varphi}
+\eta^{\tilde{S}} \frac{\partial}{\partial \tilde{S}}
+\eta^{F} \frac{\partial}{\partial F}
+\eta^{R} \frac{\partial}{\partial R}  .
\end{equation}
The coefficients of generator~(\ref{generators_generic}) depend on  
$   t $, $s$,  $ \varphi$, $S$, $F$,  $G$  and $R$ 
while the coefficients of~(\ref{generators_gamma}) 
depend on  $  t $, $s$,  $ \varphi$, $\tilde{S}$, $F$  and  $  R $.

Computations provide  the generators
\begin{multline}    \label{equivalence_general}
X ^e  _1 =  {\ddt} ,
\qquad
X ^e  _2 =  {\dds},
\qquad
X ^e  _3 =  4  t {\ddt}  
+  {\varphi} \frac{\partial}{\partial\varphi}  
+ 2 ( \gamma -2 )  S  \frac{\partial}{\partial S} 
-  F  \frac{\partial}{\partial F}  ,
\\
X ^e  _4
=   2  t {\ddt}  
-   s  {\dds} 
+ {\varphi} \frac{\partial}{\partial\varphi}   
+ 2 ( \gamma -3 )  S  \frac{\partial}{\partial S}  
+    G  \frac{\partial}{\partial G}     , 
\\
X ^e  _5
=      t {\ddt}  
-   s  {\dds} 
 +  2 ( \gamma -2 )  S  \frac{\partial}{\partial S}   
-   {R} \frac{\partial}{\partial R}    
\end{multline}
for equation~(\ref{variational_PDE}). 
For PDE~(\ref{variational_PDE_gamma}) 
we get  the generators
\begin{multline}     \label{equivalence_gamma_2}
X_{1}^{e}=  \frac{\partial}{\partial t} ,
\qquad
X_{2}^{e}=\frac{\partial}{\partial s }  ,
\qquad
X ^e  _3 =  2  t {\ddt}  
 -  2s  {\dds}  
+   {\varphi} \frac{\partial}{\partial\varphi} 
+   2  \tilde{S}  \frac{\partial}{\partial  \tilde{S}  }  ,
\\
X ^e  _4
=     2  t {\ddt}   
+  {\varphi} \frac{\partial}{\partial\varphi}  
- F  \frac{\partial}{\partial F}    , 
\qquad
X ^e  _5
=    t {\ddt}  
 +   2s  {\dds}    
+   {R} \frac{\partial}{\partial R}  , 
\\
X ^e  _6 
=  t ^2 {\ddt}  
+ t {\varphi} \frac{\partial}{\partial\varphi} 
-  t {F} \frac{\partial}{\partial F}   .
\end{multline}



\end{document}